\documentclass[a4paper,11pt]{article}
\usepackage{jheppub}
\usepackage[T1]{fontenc}
\usepackage{lmodern}
\usepackage{amstext}
\usepackage{amsfonts}
\usepackage{mathrsfs}
\usepackage{amsmath}
\usepackage{amssymb}
\usepackage{cases}
\usepackage{bm}
\usepackage{extarrows}
\usepackage{tensor}
\usepackage{xcolor}
\usepackage{float}
\usepackage{bbm, dsfont}
\usepackage{stmaryrd}
\usepackage{tocbasic}
\usepackage{slashed}
\DeclareTOCStyleEntry[
beforeskip=0.3em plus 1pt,
pagenumberformat=\textbf
]{tocline}{section}
\usepackage{subcaption}
\usepackage{graphicx}
\usepackage{extarrows}
\usepackage{tensor}
\usepackage{stmaryrd}
\usepackage{hyperref}
\usepackage{tabularx}
\usepackage{booktabs}
\usepackage{multirow}

\newcommand{\eqrefe}{Eq.\,\eqref}
\newcommand{\beqs}{\begin{subequations}}
\newcommand{\eeqs}{\end{subequations}}
\newcommand{\C}[2]{C_{{#1}}^{}\hspace*{-0.1mm}\big[\hspace*{0.1mm}{#2}\hspace*{0.1mm}\big]}

\renewcommand{\rm}{\mathrm}
\makeatletter
\newsavebox\myboxA
\newsavebox\myboxB
\newlength\mylenA
\newcommand*\xoverline[2][0.75]{%
\sbox{\myboxA}{$\m@th#2$}%
\setbox\myboxB\null
\ht\myboxB=\ht\myboxA%
\dp\myboxB=\dp\myboxA%
\wd\myboxB=#1\wd\myboxA
\sbox\myboxB{$\m@th\overline{\copy\myboxB}$}
\setlength\mylenA{\the\wd\myboxA}
\addtolength\mylenA{-\the\wd\myboxB}%
\ifdim\wd\myboxB<\wd\myboxA%
\rlap{\hskip 0.9\mylenA\raisebox{0.2ex}{\usebox\myboxB}}{\usebox\myboxA}%
\else
\hskip -0.8\mylenA\rlap{\usebox\myboxA}{\hskip 0.55\mylenA\raisebox{0.3ex}{\usebox\myboxB}}%
\fi}
\makeatother
\def\({\left(}
\def\){\right)}
\def\[{\left[\,}
\def\]{\,\right]}
\def\LB{\left\{}
\def\RB{\right\}}
\def\pp{\prime}
\def\to{\rightarrow}
\def\ito{\!\rightarrow\!}
\def\nn{\nonumber}
\def\pd{\partial}

\def\ii{\text{i}}
\def\geqq{\geqslant}
\def\leqq{\leqslant}
\def\a{\text{a}}
\def\A{\mathcal{A}}
\def\b{\text{b}}

\def\bC{\textbf{C}}
\def\tbC{\widetilde{\textbf{C}}}
\def\CC{\mathcal{C}}
\def\td{\text{d}}
\def\dd{\textbf{d}}
\def\bE{\bar{E}}
\def\EE{\mathcal{E}}
\def\f{\text{f}}
\def\ff{\textbf{f}}

\def\hg{\hat{g}}
\def\gg{\textbf{g}}
\def\tgg{\tilde{\textbf{g}}}

\def\jj{\text{j}}
\def\bk{\mathsf{k}}
\def\kk{\textbf{k}}
\def\KK{\mathcal{K}}
\def\L{\mathrm{L}}
\def\La{\mathcal{L}}
\def\MM{\mathbb{M}}
\def\M{\mathcal{M}}
\def\NN{\mathcal{N}}

\def\mO{\mathcal{O}}

\def\s{\text{s}}
\def\ss{\textbf{s}}
\def\S{\text{S}}
\def\SS{\mathcal{S}}
\def\TT{\mathcal{T}}
\def\uu{\textbf{u}}

\def\vt{\tilde{v}}
\def\vv{\textbf{v}}
\def\VV{\mathcal{V}}
\def\ww{\textbf{w}}
\def\XX{\textbf{X}}
\def\YY{\textbf{Y}}
\def\ZZ{\mathbb{Z}}
\def\al{\alpha}
\def\be{\beta}
\def\ga{\gamma}
\def\Ga{\Gamma}
\def\ka{\kappa}

\def\mn{\mu\nu}
\def\ep{\epsilon}
\def\vep{\varepsilon}
\def\lam{\lambda}
\def\hka{\hat{\kappa}}
\def\si{\sigma}
\def\tPhi{\tilde{\Phi}}

\def\slpd{\slashed{\partial}}
\def\bpsi{\bar{\psi}}
\def\bpsiI{\bar{\psi}^{(1)}}
\def\bpsiII{\bar{\psi}^{(2)}}
\def\psiI{\psi^{(1)}}
\def\psiII{\psi^{(2)}}
\def\phin{\phi_n^{}}
\def\phim{\phi_m^{}}
\def\hLn{h_n^{\mathrm{L}}}
\def\hLm{h_m^{\mathrm{L}}}
\def\vtn{\vartheta_{\!n}^{}}
\def\mod{\text{mod}}
\def\ct{c_\theta^{}}
\def\st{s_\theta^{}}

\def\ctt{c_{2\theta}^{}}
\def\stt{s_{2\theta}^{}}
\def\sz{s_0^{}}
\def\tz{t_0^{}}
\def\uz{u_0^{}}
\def\ms{m_{\text{s}}}
\def\mf{m_{\text{f}}}

\def\Msn{M_{\text{s},n}^{}}

\def\Mfn{M_{\text{f},n}^{}}
\def\Msj{M_{\text{s},\text{j}}^{}}
\def\Mn{M_n^{}}
\def\Mnn{M_n^{2}}
\def\Mm{M_m^{}}
\def\Mmm{M_m^{2}}
\def\Mj{M_\text{j}^{}}
\def\Mjj{M_\text{j}^{2}}

\def\MGn{\mathbb{M}_n^{}}
\def\MGnn{\mathbb{M}_n^2}
\def\MGm{\mathbb{M}_m^{}}

\def\MGj{\mathbb{M}_\text{j}^{}}

\def\vs{\vspace*{1mm}}
\def\hs{\hspace*{0.2mm}}
\def\hsm{\hspace*{-0.2mm}}

\allowdisplaybreaks

\title{Equivalence Theorems and Double-Copy Structure in Scattering Amplitudes of Massive Kaluza-Klein States with Matter Interactions}

\author[]{Kezhu Guo}
\emailAdd{kezhuguo2020@u.northwestern.edu}

\author[]{and Yanfeng Hang}
\emailAdd{yfhang@northwestern.edu}

\affiliation[]{Department of Physics and Astronomy,
\\
Northwestern University, Evanston, IL 60208, USA}

\abstract{
We investigate the scattering amplitudes of massive Kaluza-Klein (KK) states in compactified five-dimensional warped gauge and gravity theories. Focusing on tree-level $2\to2$ processes, we analyze the leading-order amplitudes involving bulk KK matter fields and KK gauge/gravitational Goldstone bosons.\ 
By imposing the gauge theory equivalence theorem (GAET) and the gravitational equivalence theorem (GRET) within warped KK theories, we systematically reconstruct the leading-order amplitudes for physical KK gauge bosons and gravitons, thereby circumventing the intricate energy cancellations inherent in physical amplitudes.\ Within this framework, the correspondence between GAET and GRET arises as a direct manifestation of the leading-order double-copy relation in the high-energy expansion.\ 
This connection provides a foundation for extending the BCJ double-copy construction to four-point amplitudes involving bulk KK matter fields, and further generalizes to arbitrary 
$N$-point cases, enabling a systematic derivation of the corresponding gravitational amplitudes with consistent incorporation of KK matter fields at leading order.

\vspace*{5mm}
\noindent
JHEP 09 (2025) 035~
arXiv:\,2504.05199 [hep-th] 
}

\begin{document}
\maketitle
\flushbottom

\section{Introduction}
\label{sec:1}

Scattering amplitudes are a powerful tool for exploring fundamental interactions and linking theory with experiment.\ While massless amplitudes have been widely studied, recent interest has shifted toward massive cases, particularly in models with massive gauge bosons and gravitons.\ 
Among these, Kaluza-Klein (KK) theory\,\cite{Kaluza:1921tu,Klein:1926tv} provides a self-consistent geometric mechanism for mass generation, where the additional degrees of freedom of massive KK gauge bosons and KK gravitons arise from absorbing the extra-dimensional components of their higher-dimensional counterparts.\ This ensures the conservation of physical degrees of freedom and a smooth massless limit without requiring an external Higgs field.\
This geometric mass-generation mechanism is systematically formulated in the scattering $S$-matrix through the KK equivalence theorem (ET), established for flat 5d KK gauge/gravity theories \cite{Chivukula:2001esy,Chivukula:2002ej,He:2004zr,Hang:2021fmp,Hang:2022rjp} and for warped 5d KK gauge/gravity theories \cite{Hang:2024uny,Chivukula:2023qrt} based on the Randall-Sundrum framework \cite{Randall:1999ee,Randall:1999vf}.\ 
These studies provide systematic formulations of the gauge theory equivalence theorem (GAET) and the gravitational equivalence theorem (GRET type-I/II) for warped KK gauge and gravity theories, which quantitatively connects the high-energy KK physical scattering amplitudes to those of the corresponding KK Goldstone bosons.

\vs

On the other hand, double-copy construction reveals a deep-seated connection between gravity and gauge theories, manifested in the principle that gravity can be understood as the ``square'' of the gauge theory.\
The double copy relation was first realized in (super) string theory through the Kawai-Lewellen-Tye (KLT) formula \cite{Kawai:1985xq}, which expresses massless genus-zero closed string amplitudes as products of two open string amplitudes. In the field theory limit, this leads to the tree-level double-copy relation between massless graviton and gauge boson amplitudes.\ The framework was later extended systematically to loop level with the introduction of the Bern-Carrasco-Johansson (BCJ) double copy method via color-kinematics (CK) duality \cite{Bern:2008qj,Bern:2010ue,Bern:2019prr}, which reconstructs massless graviton amplitudes from the squared amplitudes of massless gauge bosons.\ 
This duality was further generalized to include matter fields in the fundamental representation \cite{Johansson:2014zca,Johansson:2015oia,Johansson:2019dnu}.\
Recent studies have developed KLT- and BCJ-type double-copy formulations for compactified KK string theories \cite{Li:2021yfk,Gomis:2021ire} and for KK gauge and gravity theories arising from toroidal compactifications of flat 5d spacetime \cite{Hang:2021fmp,Hang:2022rjp,Li:2022rel}.\ 
Extensions of double-copy for the topologically massive Chern-Simons (CS) gauge/gravity theories were presented in Refs.\,\cite{Hang:2021oso,Hang:2023fkk,Gonzalez:2021bes}.

\vs

In this work, we study the structure of scattering amplitudes of massive KK states  in the compactified 5-dimensional warped gauge and gravity theories.\
We examine the structure of $2\to2$ leading-order amplitudes at tree level involving a pair of bulk scalar/fermion fields and two massive KK gauge/gravitational Goldstone bosons.\ By applying the GAET and GRET identities, we construct the corresponding leading order (LO) amplitudes for physical KK gauge and KK gravitons respectively.\ Another key idea of this work is to establish the GAET of KK gauge theories as the fundamental framework, from which the GRET of the corresponding KK gravity theories follows systematically via the leading-order double-copy relation in the high-energy expansion.\ Building on this foundation, we further explore the extended double-copy construction for KK scattering amplitudes in warped gauge and gravity theories.\ In particular, we extend the double-copy framework to four-point amplitudes involving massive KK gauge bosons and matter fields, constructing the corresponding KK gravity amplitudes through a proper color-kinematics correspondence at the leading energy order.\
In addition, we note that the works \cite{Johnson:2020pny,Momeni:2020hmc} have explored double-copy constructions in certain specialized KK models.\ One considered a scalar theory compactified on $\mathbb{R}^4\!\times\!S^1$, with an additional spectral condition imposed on KK mass spectrum, while another investigated a KK-inspired effective action incorporating extra global U(1) symmetries to have a particular mass relation required for double copy. However, these special constructions differ from the standard KK setup with orbifold $S^1/\mathbb{Z}_2$ compactification in our study.\ Their approaches do not overlap with our framework.

\vs

This paper is organized as follows.\ In Section\,\ref{sec:2}, we present the warped five-dimensional compactification with $S^1/\ZZ_2$ orbifold and provide a concise summary of the key results for GAET and GRET of type-I and II within the framework of warped KK gauge and gravity theories.
In Section\,\ref{sec:3}, we systematically compute the $2\to2$ tree-level amplitudes at leading order under the high-energy expansion, considering processes where a pair of bulk KK matter fields scatter into two gauge or gravitational Goldstone bosons.\ By applying the identities from GAET and GRET, we then reconstruct the corresponding leading order physical amplitudes by replacing the final-state Goldstone bosons respectively with two gauge bosons and gravitons.
Then in Section\,\ref{sec:4}, we explore the double-copy construction of four-point tree-level massive KK amplitudes within the warped gauge and gravity theories.\ We formulate the KK gravitational amplitudes involving bulk matter fields by extending the BCJ double-copy framework from the KK gauge amplitudes, both at the energy leading order.\ We also show that the extended double-copy construction applies universally to $N$-point ($N\!\geqq\!4$) amplitudes with KK matter fields at the energy leading order.\ Finally, we present our conclusions in Section\,\ref{sec:5}.\
Finally, the Appendices \ref{app:A}-\ref{app:C} present a number of analyses used for the discussions in main text.

\section{Equivalence Theorems in Warped 5d Compactification}
\label{sec:2}

In this work, we investigate the five-dimensional (5d) warped compactification within the framework of Randall-Sundrum (RS1) model\,\cite{Randall:1999ee}, wherein the Planck scale undergoes exponential suppression to generate the weak scale.\
This construction is characterized by a non-factorizable warped geometry, realized as a finite segment of AdS$_5$ spacetime.\ 
The setup consists of two 3-branes situated at the fixed points of an $S^1/\ZZ_2$ orbifold: the UV brane located at $y=0$ and the IR brane positioned at $y=\pi r_c$ with $r_c$ being the compactification radius.  

\vs

For convenience, we express the five-dimensional warped background metric in the conformal coordinates $(x^\mu, z)$ as:
\begin{equation}
\label{eq:ds2-CFz}
\td s^2 = e^{2\A(z)}\big(\eta_{\mu\nu} \td x^\mu \td x^\nu + \td z^2\big)\,.
\end{equation}
The warp factor $\A(z)$ is defined as follows\,\cite{Csaki:2004ay,Sundrum:2005jf,Rattazzi:2003ea}:
\begin{equation}
\label{eq:A(z)}
\A(z) = -\ln(1 + \bk z)\,, \qquad\qquad
\bk = (-\Lambda/6)^{\frac{1}{2}}\,,
\end{equation}
where $\bk$ defines the Anti-de Sitter (AdS) curvature scale and is of the order of Planck mass, and $\Lambda$ represents a negative bulk cosmological constant.\
In the conformal coordinate system, the interval is defined as $z \in [\hs0, L\hs]$, with $L \!=\! (e^{\bk\pi r_c}\!-\!1)/\bk$, which corresponds to the physical coordinate range $y \in [\hs0, \pi r_c\hs]$.\ 
Further, from \eqrefe{eq:ds2-CFz}, the 5d warped metric $g_{MN}^{}$ is conformally flat and can be expressed as $g_{MN}^{} \!=\! e^{2\A(z)}\eta_{MN}^{}$, where the 5d Minkowski metric $\eta_{MN}^{}$ follows the mostly-plus convention.

\vs

In Ref.\,\cite{Hang:2024uny}, we systematically formulated the gauge theory equivalence theorem (GAET) and the gravitational equivalence theorem (GRET) for warped Kaluza-Klein (KK) gauge and gravity theories within the general $R_\xi^{}$ gauge.\
These formulations build upon the geometric mechanism governing the mass generation of KK gauge bosons and KK gravitons and are established at the level of the scattering $S$-matrix through the KK equivalence theorem. This approach parallels earlier studies on flat five-dimensional KK gauge theories \cite{Chivukula:2001esy,Chivukula:2002ej,He:2004zr} and flat five-dimensional KK gravity theories \cite{Hang:2021fmp,Hang:2022rjp}.\
We summarize the key results from Ref.\,\cite{Hang:2024uny} below.

\paragraph{GAET}
The KK gauge equivalence theorem (GAET) in a compactified warped five-dimensional (5d) spacetime establishes a precise connection between the scattering amplitudes of longitudinally-polarized KK gauge bosons and those of the corresponding KK Goldstone bosons.\ The general form of the $N$-point GAET is given by
\beqs
\label{eq:5D-ETI3-ALQ}
\begin{align}
\label{eq:5D-ETI3-ALQ1}
&\TT\big[\hs A_{n_1}^{a_1\L},\cdots\!,A_{n_{N'}}^{a_{N'}\L};\tPhi\hs\big]
= \bC_\mod^{\hs n_i m_i}\,
\TT\big[\hs A_{m_1}^{a_15},\cdots\!,A_{m_{N'}}^{a_{N'}5};\tPhi\hs\big] + \TT_v^{}\,,
\\
\label{eq:KK-ET1-Tv}
& \TT_v^{} = \sum_{k=1}^{N'} \tbC_{\mod,k}^{\hs n_i m_i}\,
\TT\big[\hs v^{a_1}_{n_1},\cdots\!,v^{a_k}_{n_k}, A^{a_{k+1}5}_{m_{k+1}},\cdots\!, A^{a_{N'}5}_{m_{N'}};\tPhi\hs\big]\,,
\end{align}
\eeqs
with the 5d gauge boson $A_M^a\!=\!(A^a_\mu,A^a_5)$, the 4d longitudinally-polarized KK gauge boson $A_{n}^{a\L}\!=\!\ep^\L_\mu A_{n}^{a\mu}$, and the fifth component $A_{n}^{a5}$ identified as the would-be Goldstone boson via the geometric mass-generation mechanism arising from KK compactification\,\cite{Chivukula:2001esy,He:2004zr}.\ 
Here, we consider $N'$ external longitudinal KK gauge boson $A_{n}^{a\L}$ or KK Goldstone $A_{n}^{a5}$, while $\tPhi$ denotes the collection of all other physical states, accounting for the remaining $N\!-\!N'$ particles.\
In addition, the quantities $\bC_\mod^{\hs n_i m_i}$ and $\tbC_{\mod,k}^{\hs n_i m_i}$ are two multiplicative modification factors, taking the forms:
\beqs
\label{eq:Cmod-GA}
\begin{align}
\bC_\mod^{\hs n_i m_i} &=\bC^{\hs a_1}_{n_1m_1}\!\cdots\bC^{\hs a_{N'}}_{n_{N'} m_{N'}}
=\,\ii^{N'}\delta_{n_1m_1}^{}\!\cdots\delta_{n_{N'}m_{N'}}^{}+\mO(\rm{loop})\,,
\\[0mm]
\tbC_{\mod,k}^{\hs n_i m_i} &=\bC^{\hs a_{k+1}}_{n_{k+1}m_{k+1}}\!
\cdots\bC^{\hs a_{N'}}_{n_{N'} m_{N'}}=\,\ii^{{N'}-k}
\delta_{n_{k+1}m_{k+1}}^{}\!\cdots\delta_{n_{N'} m_{N'}}^{}+\mO(\rm{loop}) \,.
\end{align}
\eeqs
Further, in \eqrefe{eq:5D-ETI3-ALQ}, $\TT_v^{}$ stands for the residual term, which is suppressed under high energy expansion by the vector $v^{\mu}\!=\!\mO(1/E_n^{})$  contracting with external gauge bosons (cf.\,Appendix\,\ref{app:A.2}).\ 
With these ingredients, we can derive the GAET identity for high-energy scattering as:
\begin{equation}
\label{eq:KK-ET1-N}
\TT\big[\hs A_{n_1}^{a_1\L},\hs\cdots\!,A_{n_{N'}}^{a_{N'}\L};\tPhi\hs\big] \,=\, \bC_\mod^{\,n_i m_i}\,\TT\big[\hs A_{m_1}^{a_15},\cdots\!,A_{m_{N'}}^{a_{N'}5};\tPhi\hs\big]+ \mO(1/E_n^{})\,,
\end{equation}
where the multiplicative factor at tree-level will reduce to a simple form of $\bC_\mod^{\hs n_i m_i}=\ii^N\delta_{n_1m_1}\!\cdots
\delta_{n_{N'} m_{N'}}$\,.\
As shown in Ref.\,\cite{Hang:2024uny}, the GAET identity \eqref{eq:KK-ET1-N} holds non-trivially at leading order for the amplitudes with $N'\!\in\!2\ZZ$.\ While, amplitudes with ${N'}\!\in\!2\ZZ\!+\!1$, GAET identity \eqref{eq:KK-ET1-N} will take the trivial form at the leading order of high energy expansion.

\vs

In Ref.\,\cite{Hang:2024uny}, we explicitly proved the warped GAET for the fundamental three-point massive amplitudes of KK gauge bosons.\ 
We found that the GAET \eqref{eq:KK-ET1-N} manifests nontrivially in the three-point amplitude involving two longitudinal KK gauge bosons (Goldstone bosons) and one transverse KK gauge boson ($N=N'=3$), leading to the following identity:
\begin{equation}
\label{eq:GAET-3pt}
\big(M_{n_1}^2+M_{n_2}^2-M_{n_3}^2\big)\C{1}{\gg_{n_1}^{}\gg_{n_2}^{}\gg_{n_3}^{}} \,=\, 2\hs M_{n_1}^{} M_{n_2}^{}\C{1}{\tgg_{n_1}^{}\tgg_{n_2}^{}\gg_{n_3}^{}} \,,
\end{equation}
where $\gg_n^{}(z)$ and $\tgg_n^{}(z)$ are the eigenfunctions (wavefunctions) associated with the KK gauge boson $A^{a\mu}_n$ and its corresponding KK Goldstone boson $A^{a5}_n$ from the Fourier expansion.\ 
The quantities $\C{1}{\gg_{n_1}^{}\gg_{n_2}^{}\gg_{n_3}^{}}$ and $\C{1}{\tgg_{n_1}^{}\tgg_{n_2}^{}\gg_{n_3}^{}}$ appearing in \eqrefe{eq:GAET-3pt} denote the overall wavefunction (eigenfunction) coupling coefficients induced by integrating over the fifth-dimension coordinate $z$ weighted by $e^{\A(z)}$, and are defined as follows:
\beqs
\begin{align}
\label{eq:C-1}
\C{1}{\gg_{n_1}^{}\gg_{n_2}^{}\gg_{n_3}^{}} &= \frac{1}{L}\int_0^L\!\td z\,e^{\A(z)}\, \gg_{n_1}^{}(z)\,\gg_{n_2}^{}(z)\,\gg_{n_3}^{}(z)  \,,
\\
\C{1}{\tgg_{n_1}^{}\tgg_{n_2}^{}\gg_{n_3}^{}} &= \frac{1}{L}\int_0^L\!\td z\,e^{\A(z)}\, \tgg_{n_1}^{}(z)\,\tgg_{n_2}^{}(z)\,\gg_{n_3}^{}(z)  \,.
\end{align}
\eeqs
More details about the wavefunction couplings can be found in Appendix \ref{app:B.2}.\
Further, the eigenfunctions $\gg_n^{}(z)$ and $\tgg_n^{}(z)$ satisfy the boundary conditions
\begin{equation}
\label{eq:BC-YM}
\pd_z^{}\gg_n^{}(z)\big|_{z=0,L}=0\,,  \qquad~~~
\tgg_n^{}(z)\big|_{z=0,L}=0 \,,
\end{equation}
and the orthonormal conditions
\begin{equation}
\label{eq:Normalize-YM}
\frac{1}{L}\int_0^{L} \!\td z \,
e^{\A(z)}\hs\gg_n^{}(z)\hs\gg_m^{}(z) = \delta_{nm} \,,
\qquad~~
\frac{1}{L}\int_0^{L}\!\td z \,
e^{\A(z)}\hs\tgg_n^{}(z)\hs\tgg_m^{}(z) = \delta_{nm} \,.
\end{equation}
In addition, the equations of motion for $\gg_n^{}(z)$ and $\tgg_n^{}(z)$ are given by
\begin{equation}
\label{eq:EOM-gg-tgg}
(\A^\pp+\pd_z^{}) \pd_z^{}\gg_n^{}(z) = -\Mnn\,\gg_n^{}(z)\,,\qquad~~
\pd_z^{}(\A^\pp+\pd_z^{})\tgg_n^{}(z) = -\Mnn\,\tgg_n^{}(z)\hs,
\end{equation}
where they are connected via
\begin{equation}
\label{eq:EOM-gg-tgg-2}
\pd_z^{}\gg_n^{}(z) = -\Mn\,\tgg_n^{}(z)\,,\qquad~~~
(\A^\pp+\pd_z^{})\tgg_n^{}(z)=\Mn\,\gg_n^{}(z) \,.
\end{equation}
The solutions of $\gg_n^{},\tgg_n^{}$ are summarized in Eqs.\,\eqref{Aeq:gn-gtn} and \eqref{Aeq:g0}, and the KK mass $\Mn$ is determined by roots of the eigenvalue equation \eqref{Aeq:YM-Mn}.

\paragraph{GRET Type-I}
In KK gravity theory, the type-I KK gravitational equivalence theorem (GRET) establishes a connection between the scattering amplitudes of KK gravitons $h_n^{\mn}$ and the corresponding gravitational KK vector Goldstone bosons $V_n^{\mu}$, both with helicity states $\pm1$\,:
\beqs
\label{eq:KK-GRET-hV}
\begin{align}
& \M\big[\hs h^{\pm1}_{n_1},\cdots\!,h^{\pm1}_{n_{N'}};\tPhi\hs\big]=\,\bC_\mod^{V,n_i m_i}
\M\big[\hs V^{\pm1}_{m_1},\cdots\!,V_{m_{N'}}^{\pm1};\tPhi\hs\big] + \M_v^{}\,,
\\[1mm]
&\M_v^{} = \sum_{k=1}^{{N'}}\tbC_{\mod,k}^{V,n_i m_i}\,\M\big[\hs v_{n_1}^{\pm1},\cdots\!,v_{n_k}^{\pm1},V_{m_{k+1}}^{\pm1},\cdots\!,V_{m_{N'}}^{\pm1};\tPhi\hs\big]\,,
\end{align}
\eeqs
where $h_n^{\pm1}\!=\!\vep^{\pm1}_{\mn}h_n^{\mn}$ with the polarization tensors $\vep^{\pm1}_{\mn}\!=\!\!\frac{1}{\sqrt{2}}(\ep^{\pm}_{\mu}\ep^{\L}_{\nu}+\ep^{\L}_{\mu}\ep^{\pm}_{\nu})$.\ 
The off-diagonal component $V_n^{\mu}(\equiv h^{\mu5}_n)$ [cf.\,\eqrefe{eq:h-decompose}] is identified as the gravitational KK vector Goldstone boson, with $V_n^{\pm1}\!=\!V_n^{\mu}\ep_\mu^{\pm}$.\ 
The modification factors $\bC_\mod^{V,n_i m_i}$ and $\tbC_{\mod,k}^{V,n_i m_i}$ have a structural similarity to \eqrefe{eq:Cmod-GA}.\ 
At the tree level, they can be obtained by simply replacing all i with $-\ii$.\ Although discrepancies arise at the loop level, they fall outside the scope of this study and can be omitted.

\vs

Under the high energy expansion, the residual term $\M_v^{}$ is suppressed by the energy factor of $v_{\mn}^{\pm1}\!=\!\mO(1/E_n^{})$ (cf.\,Appendix\,\ref{app:A.2}).\ Hence, from \eqrefe{eq:KK-GRET-hV}, we can write down the following warped GRET identity for $h^{\pm1}_n\hs\text{-}\hs V^{\pm1}_n$ system:
\begin{equation}
\label{eq:KK-GRET-I}
\M\big[\hs h^{\pm1}_{n_1},\cdots\!,h^{\pm1}_{n_{N'}};\tPhi\hs\big]
\,=\, \bC_\mod^{V,n_i m_i}\M\big[\hs V^{\pm1}_{m_1},\cdots\!,V_{m_{N'}}^{\pm1};\tPhi\hs\big] +\mO(1/E_n^{}\text{-suppression})\,,
\end{equation}
where $\tPhi$ denotes other $N-N'$ possible physical states that interact with the KK gravitational fields.

\vs

By analyzing the three-point gravitational KK scattering amplitudes, we find that the GRET identity \eqref{eq:KK-GRET-I} leads to the following relation:
\begin{equation}
\label{eq:GRET-I-3pt}
\big(\MM_{n_1}^2+\MM_{n_2}^2-\MM_{n_3}^2\big)\C{3}{\uu_{n_1}^{}\uu_{n_2}^{}\uu_{n_3}^{}} \,=\, 2\hs \MM_{n_1}^{} \MM_{n_2}^{} \C{3}{\vv_{n_1}^{}\vv_{n_2}^{}\uu_{n_3}^{}} \,.
\end{equation}
Here, the wavefunction couplings $\C{3}{\uu_{n_1}^{}\uu_{n_2}^{}\uu_{n_3}^{}}$ and $\C{3}{\vv_{n_1}^{}\vv_{n_2}^{}\uu_{n_3}^{}}$ are defined as:
\beqs
\label{eq:C-3}
\begin{align}
\label{eq:C-3-a}
\C{3}{\uu_{n_1}^{}\uu_{n_2}^{}\uu_{n_3}^{}}&= \frac{1}{L}\int_0^L\!\td z\,e^{3\A(z)}\, \uu_{n_1}^{}(z)\,\uu_{n_2}^{}(z)\,\uu_{n_3}^{}(z)  \,,
\\
\label{eq:C-3-b}
\C{3}{\vv_{n_1}^{}\vv_{n_2}^{}\uu_{n_3}^{}}&= \frac{1}{L}\int_0^L\!\td z\,e^{3\A(z)}\, \vv_{n_1}^{}(z)\,\vv_{n_2}^{}(z)\,\uu_{n_3}^{}(z)  \,,
\end{align}
\eeqs
where $\uu_n^{}(z)$ and $\vv_n^{}(z)$ are the eigenfunctions associated with $h^{\mn}_n$ and $V^\mu_n$ respectively, obeying the boundary conditions:
\begin{equation}
\left.\pd_z^{}\uu_n^{}(z)\right|_{z=0,L}=0\,,\qquad~~~
\left.\vv_n^{}(z)\right|_{z=0,L}=0 \,,
\end{equation} 
as well as the orthonormal conditions
\begin{equation}
\label{eq:Normalize-u-v}
\frac{1}{L}\!\int_0^L\!\td z\,e^{3\A(z)}\hs\uu_n^{}(z)\uu_m^{}(z)=\delta_{nm}\,,\qquad~~
\frac{1}{L}\!\int_0^L\!\td z\,e^{3\A(z)}\hs\vv_n^{}(z)\vv_m^{}(z)=\delta_{nm}\,.
\end{equation}
Furthermore, the equations of motion for $\uu_n^{}(z)$ and $\vv_n^{}(z)$ are given as follows:\footnote{%
{N'}ote that the equation of motion for $\vv_n^{}(z)$ can be expressed in two supersymmetric ways \cite{Lim:2007fy,Lim:2008hi}.}
\beqs
\label{eq:EOM-un-vn}
\begin{align}
&(3\A^\pp+\pd_z^{})\pd_z^{}\uu_n^{}(z)= -\MGnn\,\uu_n^{}(z)\,,
\\[1mm]
&\pd_z^{}(3\A^\pp+\pd_z^{})\vv_n^{}(z)=(2\A^\pp+\pd_z^{})(\A^\pp+\pd_z^{})\vv_n^{}(z) = -\MGnn\,\vv_n^{}(z) \,,
\end{align}
\eeqs
where the solutions of $\uu_n^{}(z)$ and $\vv_n^{}(z)$ are given in Eqs.\,\eqref{Aeq:KK-un}-\eqref{Aeq:KK-vn} and \eqref{Aeq:u0}, while the KK mass eigenvalue $\MGn$ is determined by solving the \eqrefe{Aeq:GR-Mn}.

\paragraph{GRET Type-II}
The second type of GRET links the scattering amplitudes of longitudinal KK gravitons $\hLn$ to those of the corresponding KK scalar Goldstones $\phin$:
\beqs
\label{eq:KK-GRET-hphi}
\begin{align}
\label{eq:GET-ID0} 
& \M\big[\hs h^\L_{n_1},\cdots,h^\L_{n_{{N'}}};\tPhi\hs\big] 
=\,\bC_\mod^{\hs\phi,n_i m_i}\M\big[\hs\phi_{m_1}^{},\cdots\!,\phi_{m_{{N'}}}^{};\tPhi\hs\big]+\M_\Delta^{}\,,
\\[1mm]
\label{eq:RT-G}
&\M_\Delta^{}=\sum_{k=1}^{N'} \tbC_{\mod,k}^{\phi,n_i m_i}\,
\M\big[\hs\tilde{\Delta}_{n_1}, \cdots,\tilde{\Delta}_{n_k},\phi_{m_{k+1}}^{},\cdots,\phi_{m_{N'}}^{};\tPhi\hs\big] \,,
\end{align}
\eeqs
where $h_n^{\L}=\vep^{\L}_{\mn}h_n^{\mn}$ with $\vep^{\L}_{\mn}\!=\!\frac{1}{\sqrt{6}}(\ep_\mu^+\ep_\nu^-+\ep_\mu^-\ep_\nu^++2\ep_\mu^\L\ep_\nu^\L)$.\ The diagonal component $\phin(\equiv h^{55}_n)$ [cf.\,\eqrefe{eq:h-decompose}] is identified as the gravitational KK scalar Goldstone boson.\ 
The modification factors $\bC_\mod^{\hs\phi,n_i m_i}$ and $\tbC_{\mod,k}^{\phi,n_i m_i}$ in \eqrefe{eq:KK-GRET-hphi} are also similar to those of \eqrefe{eq:Cmod-GA}.\ 
At tree level, they are obtained by replacing all i with 1.\ The differences at loop level are ignored, since we focus on the tree-level analysis in the following text. 

\vs

Similarly, under the high energy expansion, the residual term $\M_\Delta^{}$ is suppressed by the factor of $\hs\tilde{\Delta}_n$ (cf.\,Appendix\,\ref{app:A.2}).\ Hence, from \eqrefe{eq:KK-GRET-hphi}, the GRET identity for $h^{\L}_n\hs\text{-}\hs\phin$ system can be expressed as follows:
\begin{equation}
\label{eq:KK-GRET-II}
\M\big[\hs h^\L_{n_1},\cdots\!, h^\L_{n_{N'}};\tPhi\hs\big] 
\,=\, \bC_\mod^{\hs\phi,n_i m_i}\,\M\big[\hs\phi_{m_1}^{},\cdots\!,\phi_{m_{N'}}^{};\tPhi\hs\big]+ \mO(1/E_n^{}\text{-suppression}) \,.
\end{equation}
At three-point level, \eqrefe{eq:KK-GRET-II} indicates the following relation:
\begin{equation}
\label{eq:GRET-II-3pt}
\Big[\big( \MM_{n_1}^2+ \MM_{n_2}^2-\MM_{n_3}^2\big)^{2} + 2\hs\MM_{n_1}^2\MM_{n_2}^2 \Big] \C{3}{\uu_{n_1}^{}\uu_{n_2}^{}\uu_{n_3}^{}}
=\, 6\,\MM_{n_1}^2\MM_{n_2}^2\hs \C{3}{\ww_{n_1}^{}\ww_{n_2}^{}\uu_{n_3}^{}} \,,
\end{equation}
where the definition of $\C{3}{\ww_{n_1}^{}\ww_{n_2}^{}\uu_{n_3}^{}}$ is analogous to that of \eqrefe{eq:C-3-a} [or \eqrefe{eq:C-3-b}], with $\uu_n^{}$ (or $\vv_n^{}$) replaced by $\ww_n^{}$.\ 
Here $\ww_n^{}$ is eigenfunction associated with the Goldstone state $\phin$, obeying the boundary and orthonormal conditions
\begin{equation}
\label{eq:EOM-normalize-w}
\left.(2\A'\!+\pd_z^{}) \ww_n^{}(z)\right|_{z=0,L}=0\,,\qquad~~
\frac{1}{L}\!\int_0^L\!\td z\,e^{3\A(z)}\hs\ww_n^{}(z)\ww_m^{}(z)=\delta_{nm}\,.
\end{equation} 
Finally, the equation of motion for $\ww_n^{}(z)$ is given by
\begin{equation}
\label{eq:EOM-wn}
(\A^\pp+\pd_z^{})(2\A^\pp+\pd_z^{})\ww_n^{}(z)=-\MGnn\,\ww_n^{}(z)\,,
\end{equation}
where the solution can be found in Eqs.\,\eqref{Aeq:KK-wn} and \eqref{Aeq:w0}.

\paragraph{GRET Type-I+II}
Finally, we consider the $N$-point amplitude including $K$ external KK gravitons with helicities $\pm1$, $N'\!-\!\!K$ longitudinal KK gravitons and $N\!-\!N'$ other physical-states $\tPhi$.\  
The corresponding GRET identity in this case requires combining \eqrefe{eq:KK-GRET-I} and \eqrefe{eq:KK-GRET-II} to formulate
\begin{align}
\label{eq:KK-GRET-I+II}
&\M\big[h^{\pm1}_{n_1},\cdots\!,h^{\pm1}_{n_K},h^\L_{n_{K+1}},\cdots\!, h^\L_{n_{N'}};\tPhi\hs\big]
\nn\\[1mm]
&= \bC_\mod^{V,n_i m_i} \bC_\mod^{\phi,n_j m_j} \M\big[\hs V^{\pm1}_{m_1},\cdots\!,V_{m_K}^{\pm1},\phi_{m_{K+1}}^{},\cdots\!,\phi_{m_{N'}}^{};\tPhi\hs\big] +\mO(1/E_n^{}\text{-suppression})\,.
\end{align}
Here, the modification factors are expanded as:
\begin{align}
\bC_\mod^{V,n_i m_i} = (-\ii)^{K}\delta_{n_1m_1}^{}\!\!\cdots\hs\delta_{n_{K}m_{K}}^{} \,, \qquad
\bC_\mod^{\phi,n_j m_j} = \delta_{n_{K+1}m_{K+1}}^{}\!\!\cdots\hs\delta_{n_{N'} m_{N'}}^{} \,,
\end{align}
where we have neglected the loop factors $\mO(\rm{loop})$, as we focus on the tree-level cases in the following discussion.
At three-point level, \eqrefe{eq:KK-GRET-I+II} indicates the following relation:
\begin{equation}
\Big[2\hs\MM_{n_3}^{4}\!-(\MM_{n_1}^2 \!+\MM_{n_2}^2)-(\MM_{n_1}^2 \!- \MM_{n_2}^2)^2\Big]\C{3}{\uu_{n_1}^{}\!\uu_{n_2}^{}\!\uu_{n_3}^{}}\!= 
6\hs\MM_{n_1}^{}\MM_{n_2}^{}\MM_{n_3}^2\C{3}{\vv_{n_1}^{}\!\vv_{n_2}^{}\!\ww_{n_3}^{}},
\end{equation}
where the wavefunction coupling $\C{3}{\vv_{n_1}^{}\vv_{n_2}^{}\ww_{n_3}^{}}$ is defined in a similar way as given in \eqrefe{eq:C-3}.

\vs

The GRET identity \eqref{eq:KK-GRET-I+II} has non-trivial contibutions at leading-order for $K\!\in\!2\ZZ$ with $K\!>\!0$. 
When $K\!=\!0$, \eqrefe{eq:KK-GRET-I+II} will reduce to \eqrefe{eq:KK-GRET-II}, which is valid for $N'\!\in\!2\ZZ$.

\section{Massive Amplitudes in Warped Gauge and Gravity Theories}
\label{sec:3}

In this section, we systematically analyze the leading-order amplitudes for the scattering of a pair of bulk KK scalars $(\varphi_n^{},\varphi_n^{\ast})$ and KK fermions  $(\psi_n^{},\bar{\psi}_n^{})$ into two KK Goldstone bosons within both the gauge and gravity theories, formulated in the Feynman-'t Hooft gauge scenario.\
Further, utilizing the GAET and GRET identities, we reconstruct the corresponding leading-order physical amplitudes, with the gauge bosons carrying longitudinal helicity states, and the gravitons possessing $\pm1$ and longitudinal helicity states.

\subsection{KK Gauge Theory}
\label{sec:3.1}

We first investigate the warped 5d Lagrangians describing the interactions between matter fields and gauge fields, along with the general procedure for compactifying the 5d theories to 4d.\ Then, we compute and analyze the tree-level $2 \ito 2$ scattering amplitudes for a pair of KK matter fields scattering into KK gauge/Goldstone fields.

\subsubsection{Interaction Lagrangian}
\label{sec:3.1.1}

The five-dimensional bulk Lagrangian for the matter fields and their interactions with gauge fields is given by
\begin{align}
\label{eq:La-gauge-matter}
\La_{\text{YM-Matter}}^{5\rm{d}} =&\, \sqrt{-g}\,\Big[\, g^{MN} (D^{}_{M,ki}\varphi_i^{})^\ast(D^{}_{N,kj}\varphi_j^{}) 
\!+\ms^2\varphi_i^{\ast}\delta_{ij}^{}\varphi_j^{} 
+\ii\bar{\si}_i^{}\,\Ga^A\EE_A^M\big(D_{M}^{}+\SS_{M}^{}\big)_{ij}\si_j^{} 
\nn\\[-.5mm]
&\hspace*{1.cm} 
+\ii\bar{\chi}_i^{}\,\Ga^A\EE_A^M\big( D_{M}^{}+\SS_{M}^{}\big)_{ij}\chi_j^{}\!-\!\mf^{}\hs\big(\bar{\si}_i^{}\delta_{ij}^{}\chi_j^{} \!+\! \bar{\chi}_i^{} \delta_{ij}^{} \si_j^{}\big)\Big]\,,
\end{align}
where $\varphi$ represents the 5d complex scalar field with bulk mass $\ms^{}$, while $\si$ and $\chi$ denote two types of 5d Dirac fermion fields with bulk mass $\mf^{}\hs$, introduced to resolve the non-chirality problem in 5d space.\ 
All matter fields transform in the fundamental representation, carrying a color index $i$. In \eqrefe{eq:La-gauge-matter}, the covariant derivative and spin connection are given by
\begin{equation}
D_{M,ij}^{}= \delta_{ij}^{}\pd_M^{} - \ii \hg A_M^{a} T^a_{ij} \,,\qquad~~
\SS_{M,ij}^{}=\frac{1}{8}\, \delta_{ij}^{}\, \tensor{\omega}{_M^A^B}\,\big[\Ga_A^{},\Ga_B^{}\big] ,
\end{equation}
where $A^a_M=(A^a_\mu,A^a_5)$ with $A^a_5$ being the Goldstone boson.\ 
In addition, $\hg$ is the 5d gauge coupling, $T^a_{ij}$ are the generators of SU($N$) gauge group and $\Ga^A$ are the 5d gamma matrices expressed as $\,\Ga^A\!=\!\big( \ga^\al,\, \ii \ga^5 \big)$ with $\ga^\al$ being the 4d gamma matrices and $\ga^5\!=\!\ii\ga^0\ga^1\ga^2\ga^3$.\ 
The 5d gamma matrices obey the Clifford algebra $\big\{ \Ga^A , \, \Ga^B \big\} \!=\!-2 \eta^{AB}$.\ 
More details about gamma matrices can be found in Appendix\,\ref{app:A}.\
Further, in the expression of spin connection, $\omega_M^{AB}$ can be entirely determined by the f\"{u}nfbein fields, $\EE_M^A$, which saitisfy the metric relation $g_{MN}^{}\!=\!\EE_M^{A}\EE_N^{B}\eta_{AB}^{}$. 
Finally, in \eqrefe{eq:La-gauge-matter}, we make the weak-field expansion of the 5d metric and retain only the leading-order terms, setting $g_{MN}^{}\!=\!e^{2\A(z)}\eta_{MN}^{}$ and $g^{MN}\!=\!e^{-2\A(z)}\eta^{MN}$, to analyze the interactions between the gauge and matter fields.

\paragraph{Scalar Sector}
For the bulk scalar fields given in the Lagrangian \eqref{eq:La-gauge-matter}, we can derive the equation of motion as
\begin{equation}
\Big[\pd^2_\mu+(3\A^\pp+\pd_z^{})\pd_z^{}-e^{2\A}\,\ms^2\,\Big]\varphi_i^{} = 0 \,,
\end{equation}
where the warp factor $\A(z)$ is defined in \eqrefe{eq:A(z)}.\ Then, we expand the 5d scalar field in terms of the eigenfunction of warped space, performing the Fourier series
\begin{equation}
\label{eq:S-expansion}
\varphi_i^{}(x,z) = \frac{1}{\sqrt{L}}\sum_{n=0}^{\infty} \varphi_{i,n}^{}(x)\,\ss_n^{}(z) \,,
\end{equation}
where the wavefunction (eigenfunction) $\ss_n^{}(z)$ obeys the following Neumann boundary condition and orthonomal relation:
\begin{equation}
\left.\pd_z^{} \ss_n^{}(z)\right|_{z\hs=\hs0,L}=0 \,,\qquad~~~
\frac{1}{L}\int_0^{L} \!\td z \,
e^{3\A(z)}\hs\ss_n^{}(z)\hs\ss_m^{}(z) = \delta_{nm}\hs.
\end{equation}
From this, we can derive the equation of motion for $\ss_n^{}(z)$ as
\begin{equation}
\label{eq:S-EOM}
(3\A^\pp+\pd_z^{})\hs\pd_z^{}\ss_n^{}(z) = \big(e^{2\A}\,\ms^2-M_{\s,n}^2\,\big)\ss_n^{}(z) \,,
\end{equation}
where the solution to $\ss_n^{}(z)$ is summarized in \eqrefe{Aeq:s0-sn} and the KK mass $\Msn$ is determined by \eqrefe{Aeq:S-Msn}.\
Further, under the KK expansions of the 5d scalar fields \eqref{eq:S-expansion}, the effective 4d KK Lagrangian can be derived by integrating over $z$ in the interval $[\hs0,L\hs]$.
The KK scalar Lagrangian in quadratic order is given by 
\begin{equation}
\La^{(2)}_{\s} = e^{3\A}\big(\,|\pd_\mu^{}\varphi_{i,n}^{}|^2 + m_{\s,n}^2|\varphi_{i,n}^{}|^2\,\big)\,,
\end{equation}
where $m_{\s,n}^{}\!=\!(e^{2\A}\,\ms^2+M_{\s,n}^2)^{\frac{1}{2}}$ represents the KK mass of the scalar field.\
In addition, for cubic and quartic Lagrangians including the interactions between the KK scalar fields and the KK gauge/Goldstone fields, refer to the Appendix\,\ref{app:C}.

\paragraph{Fermion Sector}
The challenge associated with fermion fields arises from the absence of chirality in five dimensions.\ 
Unlike in four dimensions, five-dimensional space does not allow for a $\Ga^5$ matrix (analogous to $\ga^5$ in 4d) that anti-commutes with all other $\Ga^M$.\ This restriction prevents a single 5d fermion field from yielding a chiral Standard Model (SM) fermion ($\psi_{i,0}^{}$) in 4d after imposing compactification \cite{Macesanu:2005jx}.\
Therefore, to recover the SM fermions, we introduce two 5d fermion fields in \eqrefe{eq:La-gauge-matter}: $\si_i^{}$, carrying the quantum numbers of the left-handed $\psi_{i,0}^{L}=P_L^{}\si_{i,0}^{}$ in 4d, and $\chi_i^{}$, carrying those of the right-handed $\psi_{i,0}^{R}=P_R^{}\chi_{i,0}^{}$ in 4d, where the projection operators are defined $P_{L/R}^{}\!=\!(1\pm\ga^5)/2$.\  
These two 5d fermion fields are connected via the following Dirac equations:
\beqs
\label{eq:5d-Dirac}
\begin{align}
&\big[\ii\slpd-\ga^5(2\A^\pp+\pd_z^{})\big]\si_i^{}  = e^{\A}\,\mf^{}\,\chi_i^{} \,,
\\[1mm]
&\big[\ii\slpd-\ga^5(2\A^\pp+\pd_z^{})\big]\chi_i^{} =  e^{\A}\,\mf^{}\,\si_i^{} \,.
\end{align}
\eeqs
The orbifold parity transformation $z\ito-z$ then enables the recovery of left- and right-handed fermions.\ We observe that $\ga^5$ acts as a parity operator for the fifth coordinate $z$, and require the two fermion fields satisfy the orbifold symmetry
conditions under $z$-parity, respectively \cite{Cheng:1999bg,Georgi:2000wb}:  
\begin{equation}
\label{eq:ga5-z-parity}
\ga^{5}\,\si_i^{}(x, -z) = -\si_i^{}(x,z)\,, \qquad~~~
\ga^{5}\,\chi_i^{}(x, -z)= \chi_i^{}(x,z)\,.
\end{equation} 
Thus, we can expand the two 5d fermion fields in terms of the Fourier series according to \eqrefe{eq:ga5-z-parity} and obtain:
\beqs
\label{eq:F-expansion}
\begin{align}
\si_i^{}(x,z) &= \frac{1}{\sqrt{L}} \[\sum_{n=0}^{\infty} \si^{L}_{i,n}(x)\hs\dd_n^{}(z) +\sum_{m=1}^{\infty}\si^{R}_{i,m}(x)\hs\kk_m^{}(z)\],
\\[1mm]
\chi_i^{}(x,z) &= \frac{1}{\sqrt{L}} \[\sum_{n=0}^{\infty} \chi_{i,n}^{R}(x)\hs\dd_n^{}(z) +\sum_{m=1}^{\infty}\chi_{i,m}^{L}(x)\hs\kk_m^{}(z)\],
\end{align}
\eeqs
where the wavefunctions $\dd_n^{}(z)$ and $\kk_n^{}(z)$ satisfy the boundary conditions \cite{Gherghetta:2000qt}
\begin{equation}
\big(2\A'+\pd_z^{}-e^{\A}\,\mf^{}\,\big)\hs\dd_n^{}(z)\big|_{z\hs=\hs0,L}=0 \,,\qquad~~~
\kk_n^{}(z)\big|_{z\hs=\hs0,L}=0\,,
\end{equation}
and the orthonormal relations
\begin{align}
&\frac{1}{L}\int_0^{L} \!\td z \, 
e^{4\A(z)}\hs \dd_n^{}(z)\,\dd_m^{}(z) = \delta_{nm}\,, \qquad~
\frac{1}{L}\int_0^{L}\!\td z \, 
e^{4\A(z)}\hs\kk_n^{}(z)\,\kk_m^{}(z) = \delta_{nm} \,,
\nn\\
&\frac{1}{L}\int_0^{L}\!\td z \, 
e^{4\A(z)}\hs\dd_n^{}(z)\,\kk_m^{}(z) = 0 \,.
\end{align}
Thus, after compactification, the zero-mode (SM) fermion fields $\psi_{i,0}^{}$ and the fermion fields $\si_{i,n}^{}\,,\chi_{i,n}^{}$ with KK level-$n$ are obtained as:
\begin{equation}
\psi_{i,0}^{}= \si_{i,0}^{L}+\chi_{i,0}^{R}\,,\qquad
\si_{i,n}^{}  = \si_{i,n}^{L} +\si_{i,n}^{R} \,, \qquad
\chi_{i,n}^{} = \chi_{i,n}^{L} +\chi_{i,n}^{R}\,.
\end{equation}

By substituting the expansions \eqref{eq:F-expansion} into \eqrefe{eq:5d-Dirac}, we then derive 
\beqs
\begin{align}
\big(2\A'+\pd_z-e^{\A}\,\mf^{}\big)\hs\dd_n^{}(z) &= -M_{\f,n}^{}\kk_n^{}(z)\,,
\\[1mm]
\big(2\A'+\pd_z+e^{\A}\,\mf^{}\big)\hs\kk_n^{}(z) &= M_{\f,n}^{}\dd_n^{}(z)\,,
\end{align}
\eeqs
and obtain the equations of motion for $\dd_n^{}(z)$ and $\kk_n^{}(z)$ as
\beqs
\label{eq:EOM-dn-kn}
\begin{align}
e^{-2\A}\big[\pd_z^2-e^{2\A}\hs\mf^{}\hs(\mf^{}-k)\big]e^{2\A}\dd_n^{}(z) &= -M_{\f,n}^2\dd_n^{}(z)\,,
\\[1mm]
e^{-2\A}\big[\pd_z^2-e^{2\A}\hs\mf^{}\hs(\mf^{}+k)\big]e^{2\A}\kk_n^{}(z) &= -M_{\f,n}^2\kk_n^{}(z)\,,
\end{align}
\eeqs
solutions presented in Eqs.\,\eqref{Aeq:d0-dn-kn} and \eqref{Aeq:F-Mfn}.

\vs

After imposing compactification, we can derive the effective 4d KK fermion Lagrangian. Its quadratic order takes the following form:
\begin{align}
\label{eq:L-f-(2)-1}
\La_{\f}^{(2)}  \!= e^{4\A}\bigg[\bar{\psi}_{i,0}^{}\delta_{ij}^{}(\ii\slpd-e^{\A}\hs\mf^{})\psi_{j,0}^{}\!  
+\!\sum_{n=1}^{\infty}\delta_{ij}^{}
\big(\bar{\si}_{i,n}^{},\bar{\chi}_{i,n}^{}\big)
\bigg(\!\begin{array}{cc}
\ii\slpd-\!\Mfn & -e^{\A}\hs\mf^{} \\[-.5mm] -e^{\A}\hs\mf^{} &\ii\slpd+\!\Mfn
\end{array}\!\bigg)\!
\bigg(\!\begin{array}{c}
\\[-6.5mm]
\si_{j,n}^{}\\[-1mm]\chi_{j,n}^{}
\end{array}\!\bigg)\bigg].
\end{align}
Focusing on the terms of KK fermions at level\hs-\hs$n\,(\geqq\!\!1)$ in \eqrefe{eq:L-f-(2)-1}, we find that the mass matrix contains off-diagonal components $-e^{\A}\hs\mf^{}$\,.\ 
To diagnalize the mass matrix, we introduce the following SO(2) transformation \cite{Smolyakov:2012ud}:
\begin{equation}
\label{eq:diag-trans}
\begin{pmatrix}
\si_{i,n}^{}  \\[1.mm] \chi_{i,n}^{} 
\end{pmatrix}
=\begin{pmatrix}
\cos\vtn &  \sin\vtn \\[1.mm]
\sin\vtn &  -\cos\vtn
\end{pmatrix}\!
\begin{pmatrix}
\psiI_{i,n} \\[1.mm] \ga_5^{}\psiII_{i,n}
\end{pmatrix} ,
\end{equation}
where the angle $\vtn$ is given by $\tan(2\vtn) = e^{\A}\hs\mf^{}/\Mfn$.\
With this transformation \eqref{eq:diag-trans} imposed, we bring the Lagrangian \eqref{eq:L-f-(2)-1} to the standard form in 4d
\begin{equation}
\La_{\f}^{(2)} = e^{4\A}\bigg[\bar{\psi}_{i,0}^{}\delta_{ij}^{}(\ii\slpd-e^{\A}\hs\mf^{})\psi_{j,0}^{}
+\sum_{n=1}^{\infty}\sum_{a=1}^{2} \bar{\psi}_{i,n}^{(a)}\delta_{ij}^{}(\ii\slpd- m^{}_{\f,n})\psi_{j,n}^{(a)}\bigg]\,,
\end{equation}
where $m^{}_{\f,n}=(e^{2\A}\,\mf^2+M^2_{\f,n})^{\frac{1}{2}}$.\
For the cubic and quartic Lagrangian including interaction terms between KK fermion fields and KK gauge/Goldstone fields, refer to Appendix\,\ref{app:C}.

\subsubsection{Scattering Amplitudes}
\label{sec:3.1.2}

In the KK gauge theory, we consider the tree-level $2\ito2$ scattering processes for a pair of bulk KK matter fields $\Phi_{i,n}^{}\!=\!(\varphi_{i,n}^{},\psi_{i,n}^{});\,\bar{\Phi}_{i,n}^{}\!=\!(\varphi_{i,n}^{\ast},\bar{\psi}_{i,n}^{})$\hs\footnote{
Note that starting from this section, we use $\Phi$ specifically to denote matter fields.\ 
In Section\,\ref{sec:2}, we use $\tPhi$ to represent all possible physical-state fields i.e., $\tPhi=\{\Phi,A^{a\pm}_n,h^{\pm2}_n,\cdots\}$.
} 
with $n\!\in\!\mathbb{N}\,$ into two longitudinally polarized KK gauge bosons $(A^{c\L}_m,A^{d\L}_m)$ and into two corresponding KK Goldstone bosons $(A^{c5}_m,A^{d5}_m)$, both with $m\!\in\!\ZZ^+$.\
Working in the center-of-mass frame from Appendix\,\ref{app:A.2} and expanding in the high-energy limit, the amplitudes
$\TT\big[\Phi_{j,n}^{}\bar{\Phi}_{i,n}^{}\ito A^{c\L}_m A^{d\L}_m\big]$ and $\TT\big[\Phi_{j,n}^{}\bar{\Phi}_{i,n}^{}\ito A^{c5}_m A^{d5}_m\big]$ can be written as:
\beqs
\label{eq:T-Phi-ALA5}
\begin{align}
\label{eq:T-Phi-AL} 
\TT\big[\Phi_{j,n}^{}\bar{\Phi}_{i,n}^{}\ito A^{c\L}_m A^{d\L}_m\big] &\,=\,
\TT_{2}^{\,\s/\f,\L}\bE^2 + \TT_{0}^{\,\s/\f,\L}\bE^0 + \mO(1/\bE^2)\,,
\\[1mm]
\label{eq:T-Phi-A5}
\TT\big[\Phi_{j,n}^{}\bar{\Phi}_{i,n}^{}\ito A^{c5}_m A^{d5}_m\big] &\,=\,
\TT_{0}^{\,\s/\f,5}+ \mO(1/\bE^2)\,,
\end{align}
\eeqs
where $\bE=E/\Mm$ with $\Mm$ being the KK mass for gauge fields $A^{a\mu}_m (A^{a5}_m)$ determined by \eqrefe{Aeq:YM-Mn}.\ Moreover, on the right-hand side of \eqrefe{eq:T-Phi-ALA5}, the superscripts ``$\s$'' and ``$\f\hs$'' indicate the KK scalars and KK fermions being the initial-state particles respectively, while ``$\L$'' and ``5'' indicate the longitudinal KK gauge bosons and KK Goldstone bosons being the final-state particles.

\vs

Owing to the guarantee provided by the GAET identity \eqref{eq:KK-ET1-N}, the amplitude \eqref{eq:T-Phi-AL} obeys an energy cancellation mechanism:$\,E^2\ito E^0$.\
At the energy leading order $(E^0)$, Eqs.\,\eqref{eq:T-Phi-AL}-\eqref{eq:T-Phi-A5} satisfy the relation
\begin{equation}
\label{eq:gauge-amp-LO}
\TT_{0}^{\hs\s/\f,\L} \,=\, -\hs\TT_{0}^{\hs\s/\f,5} \,,
\end{equation}
where the minus sign on the right-hand side comes from the modification factor $\bC_\mod^{nm}=\ii^2$.\
Further, the leading-order amplitudes \eqref{eq:gauge-amp-LO} can be expressed as follows:
\beqs
\label{eq:T0-sf}
\begin{align}
\label{eq:T0-sf-AL}
\TT_{0}^{\,\s/\f,\L} &= g^2\(\CC_s^{}\hs\KK_s^{\hs\s/\f,\L} + \CC_t^{}\hs\KK_t^{\hs\s/\f,\L} + \CC_u^{}\hs\KK_u^{\hs\s/\f,\L}\),
\\[1mm]
\label{eq:T0-sf-A5}
\TT_{0}^{\,\s/\f,5}
&= g^2 \(\CC_s^{}\hs\KK_s^{\hs\s/\f,5}  + \CC_t^{}\hs\KK_t^{\hs\s/\f,5} + \CC_u^{}\hs\KK_u^{\hs\s/\f,5}\),
\end{align}
\eeqs
where $g$ is the 4d gauge coupling, related to the 5d coupling $\hg$ via $g=\hg/\sqrt{L}$.\ The non-Abelian group factors $(\CC_s^{},\, \CC_t^{},\, \CC_u^{})$ are built out of the SU($N$) group structure constant $f^{abc}$ and the generators $T^a_{ij}$, defined as
\begin{equation}
\( \CC_s^{},\, \CC_t^{},\, \CC_u^{} \)=\(-\ii f^{cde}T^e_{ij}\,,\, T^c_{ik}T^d_{kj}\,,\, -T^d_{ik}T^c_{kj} \),
\end{equation}
and they obey the color Jacobi identity:
\begin{equation}
\label{eq:Cstu-Jacobi-ID}
\CC_s^{}+ \CC_t^{}+ \CC_u^{} = 0 \,.
\end{equation}

Next, we analyze the explicit form of the leading-order scattering amplitudes for different initial-state particles, considering bulk KK scalars and bulk KK fermions, separately.\ 
We begin by computing the amplitude with the final states being two KK Goldstone bosons.\ 
Then, applying the GAET identity \eqref{eq:KK-ET1-N} and the sum rules derived from the wavefunction-coupling relation \eqref{eq:GAET-3pt}, we systematically reconstruct the LO amplitude for two longitudinal KK gauge bosons as final states.

\paragraph{KK Scalars}
We first examine the scattering process $\varphi_{j,n}^{}\varphi_{i,n}^{\ast}\ito A^{c5}_m A^{d5}_m$.\ In \eqrefe{eq:T0-sf-A5}, the leading-order amplitudes for each kinematic channel are computed as:
\begin{equation}
\label{eq:S-K-A5}
\KK_s^{\hs\s,5} = \ct\hs\C{3}{\ss_n^2\hs\tgg_m^2} \,, \qquad~~~
\KK_t^{\hs\s,5} = \C{3}{\ss_n^2\hs\tgg_m^2} \,, \qquad~~~
\KK_u^{\hs\s,5} = -\C{3}{\ss_n^2\hs\tgg_m^2} \,,
\end{equation}
where $(c_{n\theta},s_{n\theta})\!=\!(\cos n\theta,\hs \sin n\theta)$
with $\theta$ being the scattering angle in the center-of-mass frame.\ 
Notice in \eqrefe{eq:S-K-A5}, the amplitudes of $t$- and $u$-channel are, in fact, derived by decomposing the results of the contact diagram.\ The explicit computations of $t$- and $u$-channel diagrams contribute only at subleading order $\mO(E^{-2})$ under the high-energy expansion, not at $\mO(E^0)$.\ Further, we have simplified the $s$-channel amplitude by imposing the following completeness relation:
\begin{equation}
\label{eq:completness-1}
\sum_{\jj=0}^\infty \C{3}{\ss_n^2\hs\gg_\jj^{}}\C{1}{\XX_m^2\hs\gg_\jj^{}} 
\,=\,\sum_{\jj=0}^\infty \Big(\C{3}{\ss_n^{}\XX_m^{}\ss_\jj^{}}\Big)^{\!2}
\,=\, \C{3}{\ss_n^2\hs\XX_m^2} \,,
\end{equation}
where $\XX\!=\!\{\gg,\tgg\}$, and this case, $\XX\!=\!\tgg$.\
Then, we can reconstruct the leading-order amplitude of the process $\,\varphi_{j,n}^{}\varphi^\ast_{i,n}\ito A^{c\L}_m A^{d\L}_m\,$ as guided by the GAET identity \eqref{eq:KK-ET1-N}.\ 
Specifically, this reconstruction is based on the following two sum rule identities:
\beqs
\label{eq:SR-s-1-2}
\begin{align}
\label{eq:SR-s-1}
\sum_{\jj=0}^{\infty}r_\jj^2\hs C_3^{}[\ss_n^2\hs\gg_\jj^{}]\hs \C{1}{\gg_m^2\gg_\jj^{}} &= 2r^2 \Big(\C{3}{\ss_n^2\hs\gg_m^2}-\C{3}{\ss_n^2\hs\tgg_m^2}\Big)\,,
\\[-.5mm]
\label{eq:SR-s-2}
\sum_{\jj=0}^{\infty}r_{\s,\jj}^2\Big(\C{3}{\ss_n^{}\gg_m^{}\ss_\jj^{}}\Big)^{\!2} &= \C{3}{\ss_n^2\hs\gg_m^2} +r^2\C{3}{\ss_n^2\hs\tgg_m^2}\,,
\end{align}
\eeqs 
where $(r_{\s,\jj}^{},\,r_\jj^{},\,r)$ denote the mass ratios
\begin{equation}
\label{eq:mass-ratios-1}
r_{\s,\jj}^{}=\Msj/\Msn\,,\qquad~~
r_\jj^{}\!=\!\Mj/\Msn\,,\qquad~~
r\!=\!\Mm/\Msn \,.
\end{equation}
We have directly proved that the above conditions \eqref{eq:SR-s-1-2} in Appdenix \ref{app:B.2}.\
By imposing the two conditions in \eqrefe{eq:SR-s-1-2}, we derive the leading-order amplitudes \eqref{eq:T0-sf-AL} for each kinematic channel:
\beqs
\label{eq:S-K-AL}
\begin{align} 
\KK_s^{\hs\s,\L} &=  -\frac{\ct}{2r^2}
\sum_{\jj=0}^{\infty}(2r^2\!-\!r_\jj^2) \C{3}{\ss_n^2\hs\gg_\jj^{}} \C{1}{\gg_m^2\hs\gg_\jj^{}}\,,
\\[1mm]
\KK_t^{\hs\s,\L} &=  \frac{1}{r^2} \sum_{\jj=0}^{\infty} (1\!-\!r_{\s,\jj}^2) \Big(\C{3}{\ss_n^{}\gg_m^{}\ss_\jj^{}}\Big)^{\!2} \,,
\\[1mm]
\KK_u^{\hs\s,\L} &=  -\frac{1}{r^2} \sum_{\jj=0}^{\infty} (1\!-\!r_{\s,\jj}^2) \Big(\C{3}{\ss_n^{}\gg_m^{}\ss_\jj^{}}\Big)^{\!2} \,,
\end{align}
\eeqs
where we have again applied the \eqrefe{eq:completness-1} in the derivations, setting $\XX=\gg$.

\paragraph{KK Fermions}
Next, we consider the process of $\psi_{j,n}^{\pm}\bar{\psi}_{i,n}^{\mp}\ito A^{c5}_m A^{d5}_m$.\ 
In \eqrefe{eq:T0-sf-A5}, for each kinematic channel, the LO amplitudes are computed
\begin{equation}
\label{eq:F-K-A5}
\KK_s^{\hs\f,5} =  -\st\hs \C{4}{\ff_n^{\,2}\hs\tgg_m^2}\,,\qquad
\KK_t^{\hs\f,5} = -\frac{\st}{1\!+\!\ct} \C{4}{\ff_n^{\,2}\hs\tgg_m^2}\,,\qquad
\KK_u^{\hs\f,5} = -\frac{\st}{1\!-\!\ct} \C{4}{\ff_n^{\,2}\hs\tgg_m^2}\,,
\end{equation}
where we have imposed the following completeness relation to simplify the $s$-channel amplitude:
\begin{equation}
\label{eq:completness-2}
\sum_{\jj=0}^\infty \C{4}{\ff_n^{\,2}\gg_\jj^{}}\C{1}{\XX_m^2\gg_\jj^{}} 
\,=\,\sum_{\jj=0}^\infty \Big(\C{4}{\ff_n^{}\XX_m^{}\ff_\jj^{}}\Big)^{\!2}
\,=\, \C{4}{\ff_n^{\,2}\XX_m^2} \,,
\end{equation}
where $\XX\!=\!\tgg$ in this case.\ 
In addition, in Eqs.\,\eqref{eq:F-K-A5}-\eqref{eq:completness-2}, the wavefunctions $\ff_0^{}(z)$ and $\ff_n^{}(z)$ are associated with the 4d SM fermion field $\psi_{i,0}^{}$ and 4d KK fermion field $\psi_{i,n}^{}\!=\!\frac{1}{\sqrt{2}}\Big[\psiI_{i,n}+\psiII_{i,n}\,\Big]$, defined as:
\begin{equation}
\ff_0^{}(z)=\dd_0^{}(z)\,,\qquad\qquad 
\ff_n^{}(z) = \frac{1}{\sqrt{2}}\big[\dd_n^{}(z) + \kk_n^{}(z) \big] \,.
\end{equation}
The reconstruction of LO amplitude $\TT\big[\psi_{j,n}^{\pm}\bar{\psi}_{i,n}^{\mp}\ito A^{c\L}_m A^{d\L}_m\big]$ is based on the following two sum rule conditions:
\beqs
\label{eq:SR-f}
\begin{align}
\label{eq:SR-f-1}
\sum_{\jj=0}^{\infty}r_\jj^2\hs \C{4}{\ff_n^{\,2}\gg_\jj^{}]C_1^{}[\gg_m^2\gg_\jj^{}} &= 2r^2 \Big(\C{4}{\ff_n^{\,2}\gg_m^2}-\C{4}{\ff_n^{\,2}\tgg_m^2}\Big)\,,
\\[-.5mm]
\label{eq:SR-f-2}
\sum_{\jj=0}^{\infty}r_{\f,\jj}^2\Big(\C{4}{\ff_n^{}\gg_m^{}\ff_\jj^{}}\Big)^{\!2} &= \C{4}{\ff_n^{\,2}\gg_m^2} + r^2\C{4}{\ff_n^{\,2}\tgg_m^2}\,,
\end{align}
\eeqs 
with the mass ratios $(r_{\f,\jj}^{},\,r_\jj^{},\,r)$ given by
\begin{equation}
\label{eq:mass-ratios-2}
r_{\f,\jj}^{}=\Msj/\Msn\,,\qquad~~
r_\jj^{}\!=\!\Mj/\Mfn\,,\qquad~~
r\!=\!\Mm/\Mfn \,.
\end{equation}
The proof of these sum rules \eqref{eq:SR-f} is similar to that of the scalar case, please refer to Appendix.\,\ref{app:B.2} for more details.
Therefore, we can derive the leading-order physical amplitudes \eqref{eq:T0-sf-AL} for each kinematic channel as
\beqs
\label{eq:F-K-AL}
\begin{align} 
\KK_s^{\hs\f,\L} &= \frac{\st}{2r^2}
\sum_{\jj=1}^{\infty}(2r^2\!-\!r_\jj^2) \C{4}{\ff_n^{\,2}\gg_\jj^{}}\C{1}{\gg_m^2\hs\gg_\jj^{}}\,,
\\[1mm]
\KK_t^{\hs\f,\L} &=  -\frac{\st}{r^2(1\!+\!\ct)} \sum_{\jj=0}^{\infty}(1\!-\!r_{\f,\jj}^2)\Big(\C{4}{\ff_n^{}\gg_m^{}\ff_\jj^{}}\Big)^{\!2}\,,
\\[1mm]
\KK_u^{\hs\f,\L} &=  -\frac{\st}{r^2(1\!-\!\ct)} \sum_{\jj=0}^{\infty}(1\!-\!r_{\f,\jj}^2)\Big(\C{4}{\ff_n^{}\gg_m^{}\ff_\jj^{}}\Big)^{\!2}\,,
\end{align}
\eeqs
with completness relation \eqref{eq:completness-2} applied, taking $\XX=\gg$.

\vs

Finally, for a possible consistency check, one can examine the flat 5d limit by taking the warped parameter $\bk\ito0\hs$.\ In this limit, the wavefunctions would take the following trigonometric forms:
\begin{align} 
\label{eq:gauge-flat-values}
\LB\gg_n^{}\,,\ss_n^{}\,,\dd_n^{}\RB = \sqrt{2}\cos\(n\pi z/L\),
\qquad\qquad
\LB\tgg_n^{}\,,\kk_n^{}\RB = \sqrt{2}\sin\(n\pi z/L\).
\end{align} 
Substituting \eqrefe{eq:gauge-flat-values} into the above amplitudes \eqref{eq:S-K-A5}, \eqref{eq:S-K-AL}, \eqref{eq:F-K-A5} and \eqref{eq:F-K-AL}, and integrating over the wavefunction couplings $\C{a}{\cdots}$ properly, one can obtain the corresponding flat-space amplitudes.

\subsection{KK Gravity Theory}
\label{sec:3.2}

The 5d bulk gravitational Lagrangian with matter fields is given by
\begin{align}
\La_{\text{GR-Matter}}^{5\rm{d}} =&\, - \sqrt{-g}\,\Big[\, g^{MN}\pd_M^{}\varphi^\ast\pd_N^{}\varphi+\ms^2|\varphi|^2 +\ii\bar{\si}\,\Ga^A\EE_A^M\big( \pd_M^{}\!+\!\SS_M^{}\big)\si 
\nn\\[0mm]
&\hspace*{1.5cm}  + \ii\bar{\chi}\,\Ga^A\EE_A^M\big( \pd_M^{}\!+\!\SS_M^{}\big)\chi
-\mf^{}\hs\big(\bar{\si}\chi + \bar{\chi}\si\big)\Big]\,,
\end{align}
where the scalar and fermion fields do not carry any color indices, and their KK expansions and the equations of motion are identical to those discussed in Section\,\ref{sec:3.1}.\ To analyze the interaction properties between the gravitational and matter fields, the weak-field expansion of the 5d metric should be retained to higher orders in 5d gravitational coupling $\hka\hs$. Specifically, we expand the 5d metric as:
\begin{equation}
g_{MN}^{}\!=e^{2\A(z)}\!\(\eta_{MN}^{}\!+\hka\hs h_{MN}^{}\),\qquad
g^{MN}\!=e^{-2\A(z)}\!\(\eta^{MN}\!\!-\hka\hs h^{MN}\!\!+\hka^2h^{MP}\tensor{h}{_P^N}\),
\end{equation}
allowing for the analysis of interactions involving one or two gravitons $h_{MN}^{}$ and the bulk matter fields.\ Moreover, the 5d graviton field $h_{MN}^{}$ is parametrized as follows:

\begin{equation}
\label{eq:h-decompose}
h_{MN}^{} = \begin{pmatrix}
h_{\mn}^{}\!-\!\frac{1}{2}\eta_{\mn}^{}\phi&~~ V_\mu^{} ~
\\[1mm]
V_\nu^{} & \phi
\end{pmatrix},
\end{equation}
where $V_{\mu}\equiv h_{\mu5}^{}=h_{5\mu}^{}$ and $\phi\equiv h_{55}^{}$ are identified as the gravitational vector and scalar Goldstone bosons, respectively. After compactification, all components give rise to 4d KK modes, $h^{\mn}_n,V^\mu_n$ and $\phin$, each associated with wavefunctions $\uu_n^{}(z),\vv_n^{}(z)$ and $\ww_n^{}(z)$.

\subsubsection{Type-I Scattering Amplitudes}

We start by considering the scattering of a pair of bulk KK matter fields $(\Phi_{n}^{},\bar{\Phi}_{n}^{})$ into two KK gravitons $h^{\mn}_m$ with helicity $\pm1$, and into two corresponding gravitational KK vector Goldstone bosons $V_m^\mu$.\
Under the high energy expansion, the physical amplitudes
$\M\big[\Phi_{n}^{}\bar{\Phi}_{n}^{} \ito h^{\pm1}_m h^{\mp1}_m\big]$ and corresponding Goldstone amplitude $\M\big[\Phi_{n}^{}\bar{\Phi}_{n}^{}\ito V_m^{\pm1}V_m^{\mp1}\big]$ can be expressed as follows:
\beqs
\label{eq:M-Phi-h1-V}
\begin{align}
\label{eq:M-Phi-h1} 
\M\big[\Phi_n\bar{\Phi}_n\ito h^{\pm1}_m h^{\mp1}_m\big] &\,=\, \M_4^{\hs\s/\f,\pm1}\bE^4 + \M_2^{\hs\s/\f,\pm1}\bE^2 + \mO(\bE^0) \,,
\\[1mm]
\label{eq:M-Phi-V}
\M\big[\Phi_n\bar{\Phi}_n\ito V^{\pm1}_m V^{\mp1}_m \big] &\,=\,
\M_2^{\hs\s/\f,V}\bE^2 + \mO(\bE^0)\,,
\end{align}
\eeqs
where $\bE=E/\MGm$ with $\MGm$ being the KK mass for gravitational fields $h^{\mn}_m (V^\mu_m)$ determined by \eqrefe{Aeq:GR-Mn}.\ As guaranteed by the GRET identity \eqref{eq:KK-GRET-I}, the physical amplitude \eqref{eq:M-Phi-h1-V} exhibits an energy cancellation process, $E^4\ito E^2$.\ 
At the energy leading order $(E^2)$, Eqs.\,\eqref{eq:M-Phi-h1}-\eqref{eq:M-Phi-V} satisfy the following relation:
\begin{equation}
\label{eq:gr-LO-amp-I}
\M_2^{\hs\s/\f,\pm1} =\, -\hs\M_2^{\hs\s/\f,V} \,,
\end{equation}
where the minus sign originates from $\bC^{V,nm}_\mod\!=\!(-\ii)^2$.\
In the following, we separately analyze scattering processes with bulk KK scalars and KK fermions as the initial states, focusing on their leading-order amplitudes.

\paragraph{KK Scalars}
For the scattering process of $\varphi_n^{}\varphi_n^{\ast}\ito V_m^{\pm1}V_m^{\mp1}$, only the $s$-channel mediated by $\jj$-mode KK gravitons provides non-trivial contribution.\ Thus, it is straightforward to obtain the LO scattering amplitude as
\begin{equation}
\label{eq:MT-S-V}
\M_2^{\hs\s,V} \,=\, -\frac{\ka^2}{32}(1\!-\!\ctt)\hs\sz\hs\C{3}{\ss_n^2\vv_m^2}\,,
\end{equation}
where $\ka=\hka/\sqrt{L}$ is the 4d gravitational coupling.\ 
Additionally, the following completness relation has been imposed in the derivation of \eqrefe{eq:MT-S-V}
\begin{equation}
\label{eq:completness-3}
\sum_{\jj=0}^\infty \C{3}{\ss_n^{\,2}\uu_\jj^{}}\C{3}{\XX_m^2\uu_\jj^{}} 
\,=\,\sum_{\jj=0}^\infty\Big(\C{3}{\ss_n^{}\XX_m^{}\ss_\jj^{}}\Big)^{\!2}
\,=\, \C{3}{\ss_n^{\,2}\XX_m^2} \,,
\end{equation}
with $\XX\!=\!\{\uu,\vv,\ww\}$, and in this case, $\XX\!=\!\uu$.\
Further, setting $(n_1,n_2,n_3)=(n,n,\jj)$ in the three-point wavefunction coupling relation \eqref{eq:GRET-I-3pt} and applying the completness relation \eqref{eq:completness-3} in amplitude \eqref{eq:MT-S-V}, we then construct the LO physical amplitude $\M_2^{\hs\s,\pm1}$ as follows:
\begin{equation}
\M_2^{\hs\s,\pm1} \,=\, 
\frac{\ka^2}{64}(1-\ctt)\sz\sum_{\jj=0}^\infty(r_\jj^2/r^2-2)\hs\C{3}{\ss_n^2\uu_\jj^{}}\C{3}{\uu_m^2\uu_\jj^{}}\,, 
\end{equation}
where the mass ratios
\begin{equation}
\label{eq:mass-ratio-s-gr}
r_\jj^{}= \MGj/\Msn \,, \qquad~~
r=\MGm/\Msn\,.
\end{equation}

\paragraph{KK Fermions}
For the scattering process of $\psi_n^{\pm}\bar{\psi}_n^{\mp}\ito V_m^{\pm1}V_m^{\mp1}$, the $s$-channel mediated by the $\jj$-mode KK gravitons, along with the non-trivial contributions from the $t$- and $u$-channel diagrams, collectively yields the following leading-order amplitude:
\begin{equation}
\label{eq:MT-F-V}
\M_2^{\hs\f,V} =\, \frac{\ka^2}{32}(\st\!-\!\stt)\hs\sz\hs\C{4}{\ff_n^{\,2}\hs\vv_m^2} \,,
\end{equation}
where the similar completness relation has been imposed
\begin{equation}
\label{eq:completness-4}
\sum_{\jj=0}^\infty \C{4}{\ff_n^{\,2}\uu_\jj^{}}\C{3}{\XX_m^2\uu_\jj^{}}
\,=\,\sum_{\jj=0}^\infty \Big(\C{4}{\ff_n^{}\XX_m^{}\ff_\jj^{}}\Big)^{\!2}
\,=\, \C{4}{\ff_n^{\,2}\XX_m^2} \,,
\end{equation}
with $\XX\!=\!\{\uu,\vv,\ww\}$, and $\XX\!=\!\uu$ in this case.\
Similar to the KK scalar case, by applying the three-point wavefunction coupling relation \eqref{eq:GRET-I-3pt} and the completness relation \eqref{eq:completness-4} in \eqrefe{eq:MT-F-V}, we can construct the LO physical amplitude
\begin{equation}
\M_{2}^{\hs\f,\pm1} \,=\, \frac{\ka^2}{64}(\st\!-\!\stt)\hs\sz\sum_{\jj=0}^\infty(r_\jj^2/r^2-2)\hs\C{3}{\ff_n^{\,2}\uu_\jj^{}}\C{3}{\uu_m^2\uu_\jj^{}}\,, 
\end{equation}
where the mass ratios are given by
\begin{equation}
\label{eq:mass-ratio-f-gr}
r_\jj^{}= \MGj/\Mfn \,, \qquad~~
r=\MGm/\Mfn\,.
\end{equation}

\subsubsection{Type-II Scattering Amplitudes}

In this section, we study the scattering of a pair of KK bulk matter fields $(\Phi_{n}^{},\bar{\Phi}_{n}^{})$ into two longitudinally-polarized KK gravitons $\hLm$ and into two corresponding gravitational KK scalar Goldstone bosons $\phim$ respectively.\ 
Under the high energy expansion, these amplitudes are given by
\beqs
\label{eq:M-Phi-hLh55}
\begin{align}
\label{eq:M-Phi-hL} 
\M\big[\Phi_n^{}\bar{\Phi}_n^{}\ito\hLm\hLm\hs\big] &\,=\, \M_6^{\hs\s/\f,\L}\bE^6 + \M_4^{\hs\s/\f,\L}\bE^4 + \M_2^{\hs\s/\f,\L}\bE^2 +  \mO(\bE^0) \,,
\\[1mm]
\label{eq:M-Phi-h55}
\M\big[\Phi_n^{}\bar{\Phi}_n^{}\ito\phim\phim\hs\big] &\,=\,
\M_2^{\hs\s/\f,\phi} \bE^2 + \mO(\bE^0)\,,
\end{align}
\eeqs
where $\bE=E/\MGm$ with $\MGm$ being the KK mass for gravitational fields $h^{\mn}_m (\phim)$ determined by \eqrefe{Aeq:GR-Mn}.\ As ensured by the GRET identity \eqref{eq:KK-GRET-II}, the amplitude \eqref{eq:M-Phi-hLh55} exhibits an energy cancellation process, $E^6\ito E^2$.\ 
At the energy leading order $(E^2)$, Eqs.\,\eqref{eq:M-Phi-hL}-\eqref{eq:M-Phi-h55} satisfy the following relation:
\begin{equation}
\label{eq:gr-amp-LO-II}
\M_2^{\hs\s/\f,\L} =\, \M_2^{\hs\s/\f,\phi} \,.
\end{equation}

\paragraph{KK Scalars}
For the scattering process of $\varphi_n^{}\varphi_n^{\ast}\ito\phim\phim$, the $s$-channel mediated by $\jj$-mode KK gravitons and the contact diagram have non-trivial contributions.\ 
Summing over the two diagrams, we obtain the following amplitude:
\begin{equation}
\label{eq:MT-S-phi}
\M_{2}^{\hs\s,\phi} \,=\, 
\frac{\ka^2}{32}(1\!-\!\ctt)\hs\sz\hs\C{3}{\ss_n^2\ww_m^2}\,,
\end{equation}
where we have imposed the completness relation \eqref{eq:completness-3} in the derivation.\
Further, taking $(n_1,n_2,n_3)\!=\!(n,n,\jj)$ in \eqrefe{eq:GRET-I-3pt} and applying the completness relation \eqref{eq:completness-3} in Goldstone amplitude \eqrefe{eq:MT-S-phi}, we then construct the physical amplitude as
\begin{equation}
\M_2^{\hs\s,\L} \,=\, \frac{\ka^2}{192}(1\!-\!\ctt)\sz\sum_{\jj=0}^\infty\big[(r_\jj^2/r^2-2)^2+2\,\big]\C{3}{\ss_n^2\uu_\jj^{}}\C{3}{\uu_m^2\uu_\jj^{}}\,,
\end{equation}
with the mass ratios $r_\jj^{}$ and $r$ given in \eqrefe{eq:mass-ratio-s-gr}.

\paragraph{KK Fermions}
For the scattering process of $\psi_n^{\pm}\bar{\psi}_n^{\mp}\ito\phim\phim$, the $s$-channel exchanging $\jj$-mode KK gravitons, along with the non-trivial contributions from the $t$- and $u$-channel diagrams, collectively yields the leading-order amplitude 
\begin{equation}
\label{eq:MT-F-phi}
\M_2^{\hs\f,\phi}\,=\,\frac{\ka^2}{32}\hs\stt\hs\sz\hs\C{4}{\ff_n^{\,2}\hs\ww_m^2} \,,
\end{equation}
where the completness relation \eqref{eq:completness-4} has been applied.\
Similar to the KK scalar case, by applying the three-point wavefunction coupling relation \eqref{eq:GRET-II-3pt} and the completness relation \eqref{eq:completness-4} in \eqrefe{eq:MT-F-phi}, we construct the LO scattering amplitude $\M_2^{\hs\f,\L}$ as follows:
\begin{equation}
\M_2^{\hs\f,\L} \,=\, \frac{\ka^2}{192}\stt\hs\sz\sum_{\jj=0}^\infty\big[(r_\jj^2/r^2-2)^2+2\,\big]\C{4}{\ff_n^{\,2}\uu_\jj^{}}\C{3}{\uu_m^2\uu_\jj^{}}\,,
\end{equation}
with the mass ratios $r_\jj^{}$ and $r$ given in \eqrefe{eq:mass-ratio-f-gr}.

\section{Extended Double-Copy Construction for KK Amplitudes}
\label{sec:4}

In this section, we explore the double-copy construction for the gravitational scattering amplitudes involving a pair of bulk KK matter fields as the initial states.\ Specifically, we examine the four-point KK graviton amplitudes and its corresponding gravitational KK Goldstone boson amplitudes at the LO of high energy expansion.

\vs

For scattering processes involving matter fields, the expected double-copy correspondences are guided by the helicity structure of the initial and final states.\
We begin by examining the initial-state matter fields.\ 
The double-copy framework then implies the following mappings: 
\vspace*{-4mm}
\beqs
\label{eq:Matter-correspond}
\begin{align}
\label{eq:s-s-s}
\varphi_{i,n}^{}\otimes\varphi_{i,n}^{} &~\to~ \varphi_n^{} \,,
\\[1mm]
\label{eq:s-f-f}
\psi_{i,n}^{} \otimes \varphi_{i,n}^{}&~\to~ \psi_n^{} \,.
\end{align}
\eeqs
These indicate that a scalar in KK gravity theory can be viewed as the two copies of the scalar state in KK gauge theory, while a fermion in KK gravity theory emerges from the combination of a fermion and a scalar in KK gauge theory.\
As for final states, we have correspondence relations \cite{Hang:2021fmp,Hang:2022rjp}:
\vspace*{-1mm}
\beqs
\label{eq:YM-GR-correspond}
\begin{align}
\label{eq:A-A-h}
A_n^{a\mu}\otimes A_n^{a\nu} &~\to~ h^{\mn}_n \,,
\\[1mm]
\label{eq:A5-A5-phi}
A_n^{a5}\otimes A_n^{a5} &~\to~ h^{55}_n(\equiv\phin) \,,
\\[1mm]
\label{eq:A-A5-V}
A_n^{a\mu}\otimes A_n^{a5} &~\to~ h^{\mu5}_n(\equiv V^{\mu}_n) \,.
\end{align}
\eeqs
In \eqrefe{eq:A-A-h}, it is instructive to note that the physical spin-2 KK graviton field $h^{\mn}_n$ arises from the double-copy of two spin-1 KK gauge fields.\ 
In Eqs.\,\eqref{eq:A5-A5-phi}-\eqref{eq:A-A5-V}, the gauge KK Goldstone boson $A_n^{a5}$ has two double-copy counterparts $\phin$ and $V^\mu_n$ corresponding to the scalar and vector gravitational KK Goldstone bosons.\ 
Further, from \eqrefe{eq:A-A-h}, we expect correspondence relations for helicity-0 and helicity-$\pm1$ states:\,$A_n^{a\L}\!\otimes\!A_n^{a\L}\!\to\!\hLn$ and $A_n^{a\L}\!\otimes\!A_n^{a\pm}\!\to\!h_n^{\pm1}$, implying that the longitudinal KK graviton emerges from the double-copy of two longitudinal KK gauge bosons, while the helicity-$\pm1$ graviton arises from the combination of one longitudinal and one transverse KK gauge boson.

\vs

The Eqs.\,\eqref{eq:Matter-correspond}-\eqref{eq:YM-GR-correspond} reflect a set of expected double-copy correspondences between KK gauge and gravity fields, including their associated matter contents.\ 
These relations arise from straightforward tensor product decompositions of single-particle states.\
However, in interacting theories, particularly those involving compactified extra dimensions and nontrivial couplings to bulk KK fields, these correspondences are not guaranteed and should be explicitly verified at the $S$-matrix level.\
Therefore, in the following discussions of this section, we compute the leading-order four-point scattering amplitudes involving the KK fields appearing in Eqs.\,\eqref{eq:Matter-correspond}-\eqref{eq:YM-GR-correspond} at tree level and further extend the analysis to the general $N$-point case.\
These results provide a nontrivial test of the consistency and applicability of the double-copy construction in an interacting, compactified framework.

\newpage
\subsection{Construction of $\Phi_n^{}\bar{\Phi}_n^{}\ito\hLm\hLm$ and $\Phi_n^{}\bar{\Phi}_n^{}\ito\phim\phim$ Amplitudes}
\label{sec:4.1}

To construct the four-point gravitational amplitudes describing the scattering of a pair of bulk KK matter fields into either two longitudinal KK gravitons $\hLn$ or two gravitational KK scalar Goldstone bosons $\phin$, we begin with the corresponding amplitudes in the KK gauge theory.\ 
In particular, we consider the amplitudes given in \eqrefe{eq:T0-sf}, which serve as the foundational input for the double-copy construction.\ 
Applying the high-energy expansion, we express the leading-order amplitudes \eqref{eq:T0-sf} in the following form:
\begin{equation}
\label{eq:T0-AL-A5}
\TT_0^{\,\s/\f,\L} = g^2 \sum_k \frac{~\CC_k^{}\,\NN_{k}^{\,\s/\f,\L}\,}{s_{0k}^{}} \,, \qquad~~~
\TT_0^{\,\s/\f,5} = g^2 \sum_k \frac{~\CC_k^{}\,\NN_{k}^{\,\s/\f,5}\,}{s_{0k}^{}} \,,
\end{equation}
where the index $k\!\in\!\{s,t,u\}$ and $s_{0k}^{}\!\in\!\{\sz,\tz,\uz\}$.\ 
The leading-order kinematic numerators 
$\NN_{k}^{\,\s/\f,\L}$ and $\NN_{k}^{\,\s/\f,5}$ are connected to the sub-amplitudes of each channel $\KK^{\,\s/\f,\L}_k$ and $\KK^{\,\s/\f,5}_k$ via the following relations:
\beqs
\label{eq:Nj-Ntj}
\begin{align}
\label{eq:Nj}
\LB\NN_s^{\,\s/\f,\L},\, \NN_t^{\,\s/\f,\L},\, \NN_u^{\,\s/\f,\L}\RB &\,=\,
\LB\sz\,\KK_s^{\hs\s/\f,\L},\, \tz\,\KK_t^{\hs\s/\f,\L},\, \uz\,\KK_u^{\hs\s/\f,\L}\RB,
\\[1mm]
\label{eq:Ntj}
\LB\NN_s^{\,\s/\f,5},\, \NN_t^{\,\s/\f,5},\, \NN_u^{\,\s/\f,5}\RB &\,=\,
\LB\sz\,\KK_s^{\hs\s/\f,5},\, \tz\,\KK_t^{\hs\s/\f,5},\, \uz\,\KK_u^{\hs\s/\f,5}\RB.
\end{align}
\eeqs
By summing over the kinematic numerators \eqref{eq:Nj} and \eqref{eq:Ntj}, we can verify that they obey the following kinematic Jacobi identities:
\begin{equation}
\sum_k \NN_k^{\,\s/\f,\L} \,=\,0\,,\qquad~~~ 
\sum_k\NN_k^{\,\s/\f,5} \,=\, 0 \,.
\end{equation}
These relations can be explicitly checked using the sub-amplitudes provided in \eqrefe{eq:S-K-A5} and \eqrefe{eq:F-K-A5}, or alternatively in \eqrefe{eq:S-K-AL} and \eqrefe{eq:F-K-AL}.\
We note that at leading order in the high-energy expansion, where KK gauge amplitudes become mass-independent and are analogous to those of 4d massless theories. 
Their kinematic structures satisfy the same Jacobi identities as in the massless case.\
The essential differences lie in the appearance of nontrivial wavefunction coupling coefficients from the compactified geometry.\
As such, our framework does not suffer from the locality issues \cite{Johnson:2020pny,Momeni:2020hmc} that arise in generic massive double-copy constructions at higher points.

\vs

Therefore, it allows us to construct the KK leading-order amplitudes of $\Phi\text{-}h_\L$ and $\Phi\text{-}\phi$ systems from \eqrefe{eq:T0-AL-A5}.\ 
Specifially, we can apply the color-kinematics (CK) duality by substituting the color factor with the corresponding kinematics numerator and switching the coupling constant \cite{Hang:2024uny}:
\begin{equation}
\label{eq:CK-4pt}
\CC_k^{} \,\to\, \NN_k^{\,\s/\f,\L}\,,\qquad
\CC_k^{} \,\to\, \NN_k^{\,\s/\f,5}\,,\qquad
g^2 \,\to\, -\frac{\ka^2}{16} \,.
\end{equation}
This yields the gravitational amplitudes involving longitudinal KK gravitons and KK Goldstone bosons with KK bulk matter fields, at leading order of $\mO(E^2)$.

\paragraph{KK Scalars}
Following the correspondence relations \eqref{eq:s-s-s}, \eqref{eq:A-A-h}, and \eqref{eq:A5-A5-phi}, and applying \eqrefe{eq:CK-4pt} to the KK scattering amplitudes \eqref{eq:T0-AL-A5} of longitudinal KK gauge bosons and KK Goldstone bosons at the leading order in the high-energy expansion, we derive the following leading-order gravitational KK amplitudes:
\beqs 
\begin{align}
&\M_{2}^{\hs\s,\L}(\rm{DC}) \,=\, -\frac{\ka^2}{16} 
\Bigg[\frac{(\NN_s^{\,\s,\L})^2}{\sz}+\frac{(\NN_t^{\,\s,\L})^2}{\tz}+\frac{(\NN_u^{\,\s,\L})^2}{\uz}\Bigg]\,,
\\
&\M_{2}^{\hs\s,\phi}(\rm{DC}) \,=\, -\frac{\ka^2}{16} 
\Bigg[\frac{(\NN_s^{\,\s,5})^2}{\sz}+\frac{(\NN_t^{\,\s,5})^2}{\tz}+\frac{(\NN_u^{\,\s,5})^2}{\uz}\Bigg]\,.
\end{align}
\eeqs 
Based on \eqrefe{eq:gr-amp-LO-II}, we can compute its either side to obtain the same leading-order amplitude.\ 
Since the forms of kinematic numerators $\NN_k^{\,\s,5}$ are much simpler than that of the $\NN_k^{\,\s,\L}$, with the GRET relation \eqref{eq:gr-amp-LO-II} we can use the kinematic numerators $\NN_k^{\,\s,5}$ to explicitly compute the leading-order gravitational amplitudes:
\begin{align}
\label{eq:MT-S-phi-DC}
\M_2^{\hs\s,\L}(\rm{DC}) = \M_2^{\hs\s,\phi}(\rm{DC})
&= -\frac{\ka^2}{16}
\Bigg[\frac{(\NN_s^{\,\s,5})^2}{\sz}+\frac{(\NN_t^{\,\s,5})^2}{\tz}+\frac{(\NN_u^{\,\s,5})^2}{\uz} \Bigg]
\nn\\
&= \frac{\ka^2}{32}(1\!-\!\ctt)\hs\sz\hs\Big(\C{3}{\ss_n^2\tgg_m^2}\Big)^{\!2}\,.
\end{align}
With these, we compare the double-copied amplitudes of KK gravitons (KK Goldstone bosons) as final-state particles
in \eqrefe{eq:MT-S-phi-DC} with the corresponding amplitudes obtained from
explicit Feynman diagram calculations in \eqrefe{eq:MT-S-phi}.\ 
We find that they have exactly the same kinematic structure except the difference 
between the two types of quartic wavefunction coupling coefficients.\
Hence, we can impose the following replacement for the 
double-copy construction of KK gauge/gravity amplitudes,
\begin{equation}
\Big(\C{3}{\ss_n^2\tgg_m^2}\Big)^{\!2}  \,\to\, \C{3}{\ss_n^2\ww_m^2} \,.
\end{equation}
This replacement effectively translates the quartic wavefunction couplings from the gauge theory to their gravitational counterparts, maintaining the expected kinematic structure.\
And its also indicates that at each KK level-$n$, we replace the mass-eigenvalue $\Mn$ of KK gauge bosons [determined by \eqrefe{Aeq:YM-Mn}] by mass-eigenvalue $\MGn$ of KK gravitons [determined by \eqrefe{Aeq:GR-Mn}], i.e., $\Mn\ito\MGn$.\
Thus, the amplitudes in KK gravity theory can be directly constructed from their gauge-theory counterparts, ensuring that the double-copy construction remains consistent.

\paragraph{KK Fermions}
Next, we consider the gravitational amplitude involving KK fermions as initial-state particles.\
Following the correspondence relations \eqref{eq:s-f-f}, \eqref{eq:A-A-h} and \eqref{eq:A5-A5-phi}, to construct the gravitational amplitude for KK fermions, we multiply the kinematic numerators $\NN^{\hs\s,5}_k$ and $\NN^{\hs\f,5}_k$ in \eqrefe{eq:T0-AL-A5} together to obtain the desired fermion initial states:
\begin{align}
\label{eq:MT-F-phi-DC}
\M_{2}^{\hs\f,\L}(\rm{DC}) = \M_{2}^{\hs\f,\phi}(\rm{DC})
&= -\frac{\ka^2}{16}\frac{1}{2}
\sum_{k\in\{s,t,u\}}\!\frac{~2\hs\NN_k^{\,\s,5}\NN^{\,\f,5}_k~}{s_{0k}^{}}
\nn\\
&=\frac{\ka^2}{32}\hs\stt\hs\sz\hs\C{3}{\ss_n^2\tgg_m^2}\C{4}{\ff_n^{\,2}\tgg_m^2}\,,
\end{align}
where $1/2$ in the first line of \eqrefe{eq:MT-F-phi-DC} is the symmetrization factor assocaited with the initial states.\
Furthermore, we impose the following replacement in the amplitude \eqref{eq:MT-F-phi-DC}:
\begin{equation}
\C{3}{\ss_n^2\tgg_m^2}\C{4}{\ff_n^{\,2}\tgg_m^2} \,\to\, \C{3}{\ff_n^{\,2}\ww_m^2} \,,
\end{equation}
which ensures the resulting amplitude reproduces the expression in \eqrefe{eq:MT-F-phi}.

\subsection{Construction of $\Phi_n^{}\bar{\Phi}_n^{}\ito h^{\pm1}_m h^{\mp1}_m$ and $\Phi_n^{}\bar{\Phi}_n^{}\ito V^{\pm1}_m V^{\mp1}_m$ Amplitudes}
\label{sec:4.2}

We now consider the scatterings of a pair of bulk KK matter fields into two KK gravitons $h^{\pm1}_n$ with helicity $\pm1$ and into two corresponding transverse KK vector Goldstone bosons $V^{\pm1}_n$.\
As shown in the relations \eqref{eq:s-s-s}, \eqref{eq:A-A-h} and \eqref{eq:A-A5-V}, the double-copy construction of gravitational amplitudes with two $V^\mu_n$ in the final state uses the scattering amplitudes
$\TT\big[\varphi_{j,n}^{}\varphi_{i,n}^\ast\ito A^{c\hs\pm}_m A^{d\hs\mp}_m\big]$ and
$\TT\big[\psi_{j,n}^{\pm}\bar\psi_{i,n}^{\mp}\ito A^{c\hs\pm}_m A^{d\hs\mp}_m\big]$ (where $A^{a\hs\pm}_m=A^{a\hs\mu}_m\ep_\mu^\pm$).\ 
The leading-order amplitudes for these processes are given by
\begin{equation}
\label{eq:T0-AT}
\TT_0^{\,\s/\f,\pm} = g^2  \Big(\CC_s^{}\KK^{\hs\s/\f,\pm}_s+\CC_t^{}\KK^{\hs\s/\f,\pm}_t +\CC_u^{}\KK^{\hs\s/\f,\pm}_u \Big) \,,
\end{equation}
where $k\!\in\!\{s,t,u\}$ and the amplitudes for each kinematic channel are given by
\beqs
\begin{align}
&\KK^{\hs\s,\pm}_s = 0 \,, \qquad\hspace*{4.5mm}
\KK^{\hs\s,\pm}_t = -(1 \!-\! \ct)\C{3}{\ss_n^2\gg_m^2} \,, \qquad\hspace*{7mm}
\KK^{\hs\s,\pm}_u = (1\!+\!\ct)\C{3}{\ss_n^2\gg_m^2} \,, 
\\[1.5mm]
&\KK^{\hs\f,\pm}_s = 0 \,, \qquad\hspace*{5mm}
\KK^{\hs\f,\pm}_t = -(1\!-\! \ct)^2s_\theta^{-1}\C{4}{\ff_n^{\,2}\gg_m^2} \,, \qquad
\KK^{\hs\f,\pm}_u = -\st\C{4}{\ff_n^{\,2}\gg_m^2} \,.
\end{align}
\eeqs
Rewriting the LO amplitude \eqref{eq:T0-AT} in the following form:
\begin{equation}
\label{eq:T0-AT-2}
\TT_0^{\,\s/\f,\pm} = g^2  \Bigg(\frac{~\CC_s^{}\,\NN_s^{\,\s/\f,\pm}\,}{\sz} +  \frac{~\CC_t^{}\,\NN_t^{\,\s/\f,\pm}\,}{\tz} +  \frac{~\CC_u^{}\,\NN_u^{\,\s/\f,\pm}\,}{\uz}\Bigg) \,,
\end{equation}
one can verify that the kinematic numerators $\NN_k^{\,\s/\f,\pm}$ obey the Jacobi identities:
\begin{equation}
\sum_k \NN_k^{\,\s,\pm} \,=\,0\,,\qquad~~~
\sum_k \NN_k^{\,\f,\pm} \,=\, 0 \,.
\end{equation}
Thus, in the $\Phi\hs\text{-}\hs h_{\pm1}$ and $\Phi\hs\text{-}\hs V_{\pm1}$ systems, the double-copy correspondence \eqref{eq:CK-4pt} extends further by incorporating the relation $\CC_k^{}\ito\NN_k^{\,\s/\f,\pm}$, enabling the systematic construction of gravitational amplitudes with final-state particles being either $h_n^{\pm1}$ or $V_n^{\pm1}$.

\paragraph{KK Scalars}
For scattering process with bulk KK scalars as initial-state particles, the double-copy construction is given by
\begin{align}
\label{eq:MT-S-V-DC}
-\M_2^{\hs\s,\pm1}(\rm{DC}) = \M_2^{\hs\s,V}(\rm{DC})
&= -\frac{\ka^2}{16}\frac{1}{2}
\sum_{k\in\{s,t,u\}}\!\frac{~2\hs\NN_k^{\,\s,\pm}\NN^{\,\s,5}_k~}{s_{0k}^{}}
\nn\\
&= \frac{\ka^2}{32}(1\!-\!\ctt)\hs\sz\hs \C{3}{\ss_n^2\gg_m^2}\C{3}{\ss_n^2\tgg_m^2} \,,
\end{align}
where we multiply the kinematic numerators $\NN^{\,\s,\pm}_k$ and $\NN^{\,\s,5}_k$ to match the helicity of the final-state Goldstone boson $V_m^{\pm1}$ and $1/2$ in the first line of \eqrefe{eq:MT-S-V-DC} is the symmetrization factor assocaited with the final Goldstone states.\

\vs

Further, similar to the case in Section\,\ref{sec:4.1}, we impose the following correspondence in the amplitude \eqref{eq:MT-S-V-DC}
\begin{equation}
\C{3}{\ss_n^2\gg_m^2}\C{3}{\ss_n^2\tgg_m^2} ~\to~
\C{3}{\ss_n^2\vv_m^2}\,,
\end{equation}
to get the form derived in \eqrefe{eq:MT-S-V}.

\paragraph{KK Fermions}
Finally, for gravitational scattering amplitudes with bulk KK fermions as the initial-state particles, we can implement the double-copy construction following Eqs.\,\eqref{eq:s-f-f}, \eqref{eq:A-A-h} and \eqref{eq:A-A5-V}:
\begin{align}
\label{eq:MT-F-V-DC}
&-\M_2^{\hs\f,\pm}(\rm{DC}) = \M_2^{\hs\f,V}(\rm{DC})= -\frac{\ka^2}{16}\frac{1}{4}
\sum_{k\in\{s,t,u\}}\!\frac{~2(\NN_k^{\,\f,\pm}\NN^{\,\s,5}_k\!+\NN_k^{\,\s,\pm}\NN^{\,\f,5}_k)~}{s_{0k}^{}}
\nn\\
&=\frac{\ka^2}{64}\LB(2\st\!-\!\stt)\C{4}{\ff_n^{\,2}\gg_m^2}\C{3}{\ss_n^2\tgg_m^2}
-\stt\hs\C{3}{\ss_n^2\gg_m^2}\C{4}{\ff_n^{\,2}\tgg_m^2}\RB,
\end{align}
where $1/4$ in the first line of \eqrefe{eq:MT-F-V-DC} is the symetrization factor associated with both initial and final states.\
Unlike the scalar case, this scenario shows a more intricate symmetry due to the presence of helicities in both the initial and final states.\ The resulting double-copy amplitude consists of two distinct contributions: the first term arises from the combination $(\psi_{i,n}^{},A_n^{a\pm})\otimes(\varphi_{i,n}^{},A_n^{a5})$, while the second term results from  $(\psi_{i,n}^{},A_n^{a5})\otimes(\varphi_{i,n}^{},A_n^{a\pm})$, which corresponds to an exchange of helicity states in the final particles of the first term.\ 
We impose the following correspondence relations for the wavefunction couplings in the two terms:
\begin{equation}
\C{4}{\ff_n^{\,2}\gg_m^2}\C{3}{\ss_n^2\tgg_m^2} \,\to\,
\C{4}{\ff_n^{\,2}\vv_m^2} \,, \qquad
\C{3}{\ss_n^2\gg_m^2}\C{4}{\ff_n^{\,2}\tgg_m^2}\,\to\,
\C{4}{\ff_n^{\,2}\vv_m^2} \,.
\end{equation}
With these substitutions, the double-copy amplitude exactly equals \eqrefe{eq:MT-F-V}.

\vs

Finally, to complete our discussion of the double-copy correspondence, we conclude this section with two additional examples.\ 
The cases illustrated above show that the amplitudes with final states $\hLn$ and $h_n^{\pm1}$ can be inferred from the amplitudes of the corresponding Goldstone bosons $\phin$ and $V_n^{\pm1}$ due to the presence of GRET identities.\
While, for the final-state KK graviton with helicities $\pm2$, the amplitudes $\M\big[\Phi_n\bar{\Phi}_n \to h^{\pm2}_m h^{\mp2}_m\big]$ at the leading order can be constructed by using Eqs.\,\eqref{eq:T0-AT}-\eqref{eq:T0-AT-2} as follows:
\begin{align}
\M_2^{\hs\s,\pm2}(\rm{DC}) \,&=\, -\frac{\ka^2}{16} 
\Bigg[\frac{(\NN_s^{\,\s,\pm})^2}{\sz}+\frac{(\NN_t^{\,\s,\pm})^2}{\tz}+\frac{(\NN_u^{\,\s,\pm})^2}{\uz}\Bigg]
\nn\\
&= \frac{\ka^2}{32} (1\!-\!\ctt)\sz\Big(\C{3}{\ss_n^2\gg_m^2}\Big)^{\!2}\,,
\\[1mm]
\M_2^{\hs\f,\pm2}(\rm{DC}) \,&=\, -\frac{\ka^2}{16} 
\Bigg(\frac{\NN_s^{\,\s,\pm}\NN_s^{\,\f,\pm}}{\sz}+\frac{\NN_t^{\,\s,\pm}\NN_t^{\,\f,\pm}}{\tz}+\frac{\NN_u^{\,\s,\pm}\NN_u^{\,\f,\pm}}{\uz}\Bigg)
\nn\\
&= -\frac{\ka^2}{32} (2\st\!-\!\stt)\sz\hs\C{3}{\ss_n^2\gg_m^2}\C{4}{\ff_n^{\,2}\gg_m^2} \,,
\end{align}
where the final-state KK gravitons correspondence follows from \eqrefe{eq:A-A-h}, namely $A_n^{a\pm}\otimes A_n^{a\pm}\ito h^{\pm2}_n$. By applying the replacements:
\begin{equation}
\Big(\C{3}{\ss_n^2\gg_m^2}\Big)^{\!2} \to \C{3}{\ss_n^2\uu_m^2}\,,\qquad~~~
\C{3}{\ss_n^2\gg_m^2}\C{4}{\ff_n^{\,2}\gg_m^2} \to \C{3}{\ff_n^{\,2}\uu_m^2}\,,
\end{equation}
we can reproduce the leading-order amplitudes $\M_2^{\hs\s,\pm2}$ and $\M_2^{\hs\f,\pm2}$ which are obtained from direct Feynman diagram calculations.\ In summary, this section has presented a comprehensive analysis of all possible double-copy correspondences with KK matter fields in the initial state and KK gravitons with all possible helicities or their associated KK Goldstone modes in the final state.

\subsection[\texorpdfstring{Leading-Order Double Copy with $N\geqq4$}{Leading-order Double Copy with \(N\geqq 4\)}]
{\boldmath Leading-order Double Copy with $N\geqq4$}
\label{sec:4.3}

So far, we have provided an explicit demonstration that the warped double-copy framework can be successfully applied to four-point KK scattering amplitudes at leading order in the high-energy expansion.\ In this subsection, we further extend this construction to general $N$-point $(N\!\geqq\!4)$ KK gauge and gravity amplitudes involving bulk matter fields.

\vs

We consider the general case of an $N$-point KK gauge scattering amplitudes at leading order, involving $\ell$-pair KK matter fields with $0\!\leqq\!2\ell\!<\!N$.\ Among these, $p$ of them are KK scalar pairs $(\varphi_{i,n}^{},\varphi_{i,n}^\ast)$ and the remining $\ell-p$ of them are the KK fermion pairs $(\psi_{i,n}^{},\bar{\psi}_{i,n}^{})$.\ 
The other $N-2\ell$ external states consist $r$ KK Goldstone fields $A^{a5}_n$ and $N-2\ell-r$ transverse KK gauge fields $A^{a\pm}_n$.\
The leading energy scaling of such $N$-point amplitude is consistent with the result established in Ref.\,\cite{Hang:2024uny}, and is given by
\begin{equation}
\label{eq:E-counting-YM}
D_E^{}\big[N\big(\varphi_{i,n}^{},\varphi_{i,n}^{\ast},\psi_{i,n}^{},\bar{\psi}_{i,n}^{},A^{a\pm}_n,A_n^{a5}\big)\big] =\, 
4-N- N_v^{} - \overline{\VV}^{\hs(3)}_{\hsm\rm{min}}\,,
\end{equation}
where the quantity $\overline{\VV}^{\hs(3)}_{\hsm\rm{min}}$ denotes the minimal number of non-derivative cubic vertices involving in the amplitude, where $\overline{\VV}^{\hs(3)}_{\hsm\rm{min}}\!=\!0\,(1)$ for $N'\!=\!\rm{even}\,\rm{(odd)}$ with $N'$ being the number of Goldstone boson $A^{a5}_n$.\ 
In addition, $N_v^{}$ represents the number of possible external state $v_\mu^{}A^{a\mu}_n$\!
\footnote{The state $v_\mu^{}A^{a\mu}_n$ can be considered as a part of $A_n^{a5}$.\ Within the general representation $N'A_n^{a5}$, one may include $N_v^{}$ states of $v_\mu^{}A^{a\mu}_n$.
}.\
In this analysis, we only focus on the KK gauge amplitude having even number of external $A^{a5}_n$ states ($N'\!\!\in\!\!2\ZZ$).\
Amplitudes with odd-point $A^{a5}_n$ generally contain mass-dependent contributions and lack a systematic double-copy realization in the KK gauge/gravity framework \cite{Hang:2024uny}.\
Therefore, the leading energy-power dependence \eqref{eq:E-counting-YM} can be simplied to
\begin{equation}
D_E^{}\big[N\big(\varphi_{i,n}^{},\varphi_{i,n}^{\ast},\psi_{i,n}^{},\bar{\psi}_{i,n}^{},A^{a\pm}_n,A_n^{a5}\big)\big] =\, 4\!-\!N  \,,
\end{equation}
with $\overline{\VV}^{\hs(3)}_{\hsm\rm{min}} = N_v^{}= 0$\,\footnote{
In the case of an odd number of external $A_n^{a5}$ states, $\overline{\VV}^{\hs(3)}_{\hsm\rm{min}}$ takes the form $\VV[A_n^{a\mu} A_m^{b\nu} A_\ell^{c5}]$ with one leg being external Goldstone boson.\ 
Its energy dependence is same as the vertex $\VV[A_n^{a\mu} A_m^{b\nu}  A_\ell^{c\rho}]v_\rho^{}$ from residue term, where one leg is contracted with external vector $v_\rho^{}=\mO(E^{-1})$.\ As a result, $N_v^{}\neq0$.}.

\vs

Hence, under these considerations, we can express the leading-order KK gauge amplitude in the following form:
\begin{align}
\label{eq:T0-N}
&\TT_{4-N}^{}\Big[\underbrace{\varphi_{i,n}^{},\varphi^\ast_{i,n}}_{p\text{-pair}},
\underbrace{\psi_{i,n}^{},\bar{\psi}_{i,n}^{}}_{(\ell-p)\text{-pair}};
\underbrace{A^{a5}_n}_r,
\underbrace{A^{a\pm}_n}_{N-2\ell-r}\Big]
\nn\\
&= g^{N-2} \C{3\ell-p+1}{2p\hs\ss_n^{},2(\ell\!-\!p)\ff_n^{},r\tgg_n^{},(N\!-\!2\ell\!-\!r)\gg_n^{}}
\nn\\
&\quad\times\!\sum_{k=1}^{\aleph} \frac{\CC_k^{}}{\mathcal{D}_k^{}}
\NN_k^{}\big[
2p\hs\varphi_{i,n}^{},2(\ell\!-\!p)\psi_{i,n}^{};rA^{a5}_n,(N\!-\!2\ell\!-\!r)A^{a\pm}_n
\big]\,,
\end{align}
where $p\in\ZZ$ and $r\in2\ZZ$.\ Further, in \eqrefe{eq:T0-N}, the total number of trivalent diagrams is given by $\aleph \!=\! \frac{(2N-5)!!}{(2\ell-1)!!}$\,\cite{Johansson:2015oia}, and $\mathcal{D}_k^{}$ denotes the product of propagator denominators.\ In \eqrefe{eq:T0-N}, the color and KK indices associated with each field or wavefunction may differ.\ For notational convenience, we have uniformly denoted them by $i,a$ and $n$.\ 
Besides, the notations $2p\hs\hs\varphi_{i,n}^{}$ or $2p\hs\hs\ss_n^{}$ indicates the total number of insertions and wavefunctions of scalar type.\ The same labeling convention applies to other KK fields and their corresponding wavefunctions.

\vs

As for the KK gravity theory, based on the double-copy correspondence relations Eqs.\,\eqref{eq:Matter-correspond}-\eqref{eq:YM-GR-correspond}, the extended CK duality with coupling replacement \cite{Hang:2024uny}:
\begin{equation}
\CC_k^{} \to \NN_k^{}\,,\qquad~~~
g^{N-2}\to-\(\frac{\ka}{4}\)^{\!\!N-2} ,
\end{equation}
we can construct the $N$-point LO gravitational amplitude from \eqrefe{eq:T0-N}
\begin{align}
\label{eq:N-dc}
&\M^{(\rm{DC})}_2\Big[\hs
\underbrace{\varphi_n^{},\varphi^\ast_n}_{q\text{-pair}},
\underbrace{\psi_n^{},\bar{\psi}_n^{}}_{(\ell-q)\text{-pair}};
\underbrace{\phin}_w\,,\!
\underbrace{V^{\pm1}_n}_{N-2\ell-w-s}\!,\,
\underbrace{h^{\pm2}_n}_s \hs\Big] 
\nn\\
=&-\(\frac{\ka}{4}\)^{\!\!N-2}
\(\!\frac{1}{\sqrt{2}}\)^{\!\!N-2q-w-s}
\sum_{k=1}^{\aleph}\sum_{\b=0}^{\ell-q}
\sum_{\a=0}^{\binom{2(\ell-q)}{2\b}} 
\sum_{\b'=0}^{N\!-2\ell-w-s}
\sum_{\a'=0}^{\binom{N\!-2\ell-w-s}{\b'}} 
\nn\\
&\C{3\ell-q-\b+1}{2(q\!+\!\b)\ss_n^{},
2(\ell\!-\!q\!-\!\b)\ff_n^{},
(s\!+\!\b')\gg_n^{},
(N\!-\!2\ell\!-\!s\!-\!\b')\tgg_n^{}}
\nn\\
\times\hs&\C{2\ell+\b+1}{2(\ell\!-\!\b)\ss_n^{},
2\b\ff_n^{},
(w\!+\!\b')\tgg_n^{},
(N\!-\!2\ell\!-\!w\!-\!\b')\gg_n^{}}
\frac{1}{\mathcal{D}_k^{}}\Big\{
\nn\\
&\NN_{k,\a,\a'}^{}\!\big[2q\hs\varphi_{i,n}^{}, 
2\b\varphi_{i,n}^{}, 
2(\ell\!-\!q\!-\!\b)\psi_{i,n}^{};
wA_n^{a5},(N\!-\!2\ell\!-\!w\!-\!s\!-\!\b')A_n^{a5},
\b'\!A_n^{a\pm}, sA_n^{a\pm} \big] \, 
\nn\\
\times\hs&\NN_{k,\a,\a'}^{}\!\big[2q\hs\varphi_{i,n}^{}\hsm,
2(\ell\!-\!q\!-\!\b)\varphi_{i,n}^{}\hsm,
2\b\psi_{i,n}^{};
w\hsm A_n^{a5},\b'\!A_n^{a5}\hsm,
(N\!-\!2\ell\!-\!w\!-\!s\!-\!\b')A_n^{a\pm}\hsm,
sA_n^{a\pm} \big]\!\Big\},
\end{align}
where the amplitude \eqref{eq:N-dc} contains $\ell$-pair KK matter fields, among which $q$ of them are KK scalar pairs and $\ell-q$ KK fermion pairs.\ The remining $N-2\ell$ external states consist $w$ KK gravitional scalar Goldstone bosons, $s$ gravitons with helicity $\pm2$ and $N-2\ell-w-s$ KK gravitational vector bosons with helicity $\pm1$.\ Here the numbers $(q,w,s)\!\in\!\mathbb{N}$.\
Moreover, the leading energy-power dependence of the KK gravitational amplitude \eqref{eq:N-dc} is $E^2$, consistent with the general power-counting rule \cite{Hang:2021fmp,Hang:2024uny}:
\begin{equation}
\label{eq:E-counting-GR}
D_E^{}\big[N\big(\varphi_n^{},\varphi_n^{\ast},\psi_n^{},\bar{\psi}_n^{},h^{\pm2}_n,V_n^{\pm1},\phin\big)\big] =\, 2(N_L^{}+1)+\sum_j \VV_j^{}\!\(N_{d_j^{}}-2+\frac{1}{2}N_{f_j^{}}\),
\end{equation}
where $N_L^{}\geqq0$ represents the number of loops and $\VV_j^{}$ denotes the number of type-$j$ vertices.\ In addition, each vertex of type-$j$ contains $N_{d_j^{}}$ partial derivatives and $N_{f_j^{}}$ fermionic legs.\  
In this work, we focus on the tree-level cases ($N_L^{}=0$), where two types of vertices $\VV_j^{}$ appear: (i).\,purely gravitational vertices with $(N_{d_j^{}},N_{f_j^{}})=(2,0)$, since each carries two partial derivatives;
(ii).\,vertices invovling a fermion pair with $(N_{d_j^{}},N_{f_j^{}})=(1,2)$, as each contains one partial derivative.\ 
Accordingly, the leading energy dependence \eqref{eq:E-counting-GR} simplifies to the following form:
\begin{equation}
D_E^{}\big[N\big(\varphi_n^{},\varphi_n^{\ast},\psi_n^{},\bar{\psi}_n^{},h^{\pm2}_n,V_n^{\pm1},\phin\big)\big] =\, 2\,,
\end{equation}
which reproduces the correct power counting in \eqrefe{eq:N-dc}.

\vs

Then, to match the structure of the leading-order amplitudes obtained from Feynman diagram computations, we must replace the products of wavefunction coefficients in the KK gauge theory with the corresponding universal KK gravitational wavefunction coefficient.\ 
The replacement takes the following form:
\begin{align}
&\C{3\ell-q-\b+1}{\cdots}\,
\C{2\ell+\b+1}{\cdots}
\nn\\[1mm]
&\to\C{\ell-q+3}{2\hs q\hs\ss_n^{},2(\ell\!-\!q)\hs\ff_n^{}, 
w\hs\ww_n^{}, (N\!-\!2\ell\!-\!w\!-\!s)\hs\vv_n^{},s\hs\uu_n^{}}\,,
\end{align}
where the wavefunctions $\{\ss_n^{},\ff_n^{},\uu_n^{},\vv_n^{},\ww_n^{}\}$ are presented in the Sections\,\ref{sec:2}-\ref{sec:3} and Appendix\,\ref{app:B.1}.\ This replacement also reflects the KK mass correspondence at each KK level-$n$ via $\Mn \ito \MGn$.\
Importantly, the validity of prescription for general $N$-point amplitudes has been explicitly checked.\ 
In our upcoming work \cite{5dBCFW}, we prove the consistency of the $N$-point KK gauge and gravity amplitudes using helicity spinor formalism in compactified higher-dimensional spacetimes and Britto-Cachazo-Feng-Witten (BCFW) recursion relations \cite{Britto:2004ap,Britto:2005fq}.

\section{Conclusion}
\label{sec:5}
 
In this work, we investigated the structure of scattering amplitudes of massive Kaluza-Klein (KK) states in compactified five-dimensional warped gauge and gravity theories.\ Specifically, we explored the key properties of the equivalence theorems for KK gauge and KK gravity theories (GAET and GRET) in the scattering processes involving additional bulk matter (scalar and fermion) fields.\ Further, we extended the double-copy construction to incorporate massive KK matter states, providing a systematic framework for constructing KK gravitational leading-order amplitudes from their KK gauge counterparts.  

\vs

In Section\,\ref{sec:2}, we summarized the essential results of GAET and GRET formulations within the general $R_\xi$
gauge framework, extending up to loop level \cite{Hang:2024uny}. 
The GAET is formulated in Eqs.\,\eqref{eq:5D-ETI3-ALQ}-\eqref{eq:KK-ET1-N}, while the GRET are presented in Eqs.\,\eqref{eq:KK-GRET-hV}-\eqref{eq:KK-GRET-I} for helicity-1 KK gravitons (type-I), and in Eqs.\,\eqref{eq:KK-GRET-hphi}-\eqref{eq:KK-GRET-II} for helicity-0 (longtitudinal) KK gravitons (type-II).
In Section\,\ref{sec:3}, we systematically analyzed the leading-order amplitudes of $2\ito2$ tree-level scattering processes involving bulk KK scalar and fermion fields alongside KK gauge and gravitational Goldstone bosons.\  
By applying the identites of GAET and GRET, we reconstructed the leading-order scattering amplitudes invovling longitudinal KK gauge bosons, longitudinal KK gravitons and helicity-$\pm1$ KK gravitons from the corresponding KK Goldstone amplitudes.\
It is important to emphasize that these replacements are valid only at leading order in the high-energy expansion; at subleading orders, mass-dependent residual terms arise and the replacements no longer apply.\
In Section\,\ref{sec:4}, building on this foundation, we extended the double-copy construction to encompass massive KK matter fields, demonstrating a consistent correspondence between KK gauge and gravity amplitudes at the four-point tree level.\
Furthermore, our analysis establishes that this warped double-copy framework remains valid at leading order for general $N$-point KK gauge and gravity amplitudes involving matter fields.\ 
At leading order, the KK amplitudes become mass-independent and closely resemble their 4d massless counterparts, enabling a consistent application of the double-copy framework with all locality issues avoided.

\vs

Our findings in this work underscore the utility of scattering amplitudes as a powerful probe of the underlying structure of higher-dimensional field theories. Through the double-copy correspondence, we have gained new insights into the mass generation mechanisms, the organization of fundamental interactions, and the intricate interplay between gauge and gravitational sectors in KK compactifications with matter fields involved.\ 
Looking forward, it would be worth extending the correspondence relations in \eqref{eq:Matter-correspond} to include purely fermionic building blocks.\ For instance, one may consider possible double-copy relations:
\begin{equation}
\psi_{i,n}^{\pm} \otimes \psi_{i,n}^{\pm} ~\to ~ A_n^{\pm}\,, \qquad\qquad
\psi_{i,n}^{\pm} \otimes \psi_{i,n}^{\mp}  ~\to ~ \varphi_n^{} \,,
\end{equation} 
where the double-copy counterparts of fermion states in the KK gauge theory give rise to the of U(1) gauge bosons and scalar fields in the corresponding KK gravity theory.\
Moreover, from a broader perspective, it is worth investigating whether the KLT double copy structure persists in superstring amplitudes compactified on internal manifolds with nontrivial KK spectra.\ Such an extension would bridge field-theoretic constructions with string-theoretic origins and shed light on the ultraviolet completion of massive double-copy structures.

\begin{acknowledgments}
We sincerely thank Hong-Jian He, Yin-Long Qiu and Wei-Wei Zhao for their insightful
discussions and valuable suggestions.\ K.G. would also like to thank A. Edison and J.J. Carrasco for their kind mentorings during the last two years.\ We acknowledge the support of
Northwestern University Amplitudes and Insight group, Department of Physics and Astronomy, and Weinberg College.
\end{acknowledgments}

\newpage
\appendix

\section{Conventions}
\label{app:A}

\subsection{Table of Notations Used in the Main Text}
\label{app:A.1}

We summarize the notations used in the main paper in Table\,\ref{tab:1}.
\begin{table}[H]
\centering
\renewcommand{\arraystretch}{1.1}
\begin{tabular}{l|l|l}
\hline
\multirow{8}{*}{Gauge}
& $\TT[\cdots]$ &Amplitudes involving matter and guage fields
\\
\cline{2-3}
&$A_M^a\!=\!(A^a_\mu,A^a_5)$ & 5d massless gauge field
\\
\cline{2-3}
&$A_n^{a\pm}\!=\!\ep^\pm_\mu A_{n}^{a\mu}$ 
&KK gauge field with helicity $\pm$
\\
\cline{2-3}
&$A_n^{a\L}\!=\!\ep^\L_\mu A_n^{a\mu}$ 
&KK gauge field with helicity 0 (longitudinal)
\\
\cline{2-3}
&$A^{a5}_n$ & KK Goldstone field
\\
\cline{2-3}
&$\gg_n^{}(z)$ & KK eigenfunction associated with $A^{a\mu}_n$ 
\\
\cline{2-3}
&$\tgg_n^{}(z)$ & KK eigenfunction associated with $A^{a5}_n$  
\\
\cline{2-3}
&$\Mn$ & KK mass for gauge fields ($A^{a\mu}_n,A^{a5}_n$)
\\ \hline
\multirow{11}{*}{Gravity} 
& $\M[\cdots]$ & Amplitudes involving matter and gravitational fields
\\
\cline{2-3}
&$h^{MN}\!\!=\!\begin{pmatrix}
h^{\mn}\!\!-\!\frac{1}{2}\phi&V^\mu\\[-.5mm]
V^\nu&\phi
\end{pmatrix}\hspace*{-2.5mm}$ & 5d massless graviton
\\
\cline{2-3}
&$h_n^{\pm2}\,(=\!\vep^{\pm2}_{\mn}h_n^{\mn})$ & KK graviton with helicity $\pm2$ 
\\
\cline{2-3}
&$h_n^{\pm1}\,(=\!\vep^{\pm1}_{\mn}h_n^{\mn})$ & KK graviton with helicity $\pm1$ 
\\
\cline{2-3}
&$h_n^{\L}\,(=\!\vep^\L_{\mn}h_n^{\mn})$ & KK graviton field with  helicity 0 (longitudinal)
\\
\cline{2-3}
&$V_n^{\pm1}\,(=\!\ep^{\pm}_{\mu}V_n^{\mu})$ & KK gravitational vector Goldstone with helicity $\pm1$
\\
\cline{2-3}
&$\phin$ & KK gravitational scalar Goldstone
\\
\cline{2-3}
&$\uu_n^{}(z)$ & KK eigenfunction associated with $h^{\mn}_n$ 
\\
\cline{2-3}
&$\vv_n^{}(z)$ & KK eigenfunction associated with $V^\mu_n$ 
\\
\cline{2-3}
&$\ww_n^{}(z)$ & KK eigenfunction associated with $\phin$ 
\\
\cline{2-3}
&$\MGn$ & KK mass for gravitational fields ($h^{\mn}_n,V^\mu_n,\phin$)
\\ \hline
\multirow{9}{*}{Matter} 
&$\Phi$ & A set of bulk KK matter fields
\\
\cline{2-3}
&$\varphi_n^{}/\varphi_n^{\ast}$ & KK scalar/anti-scalar field 
\\
\cline{2-3}
&$\psi_n^{}/\bar{\psi}_n^{}$ & KK fermion/anti-fermion field 
\\
\cline{2-3}
&$\ss_n^{}(z)$ & KK eigenfunction associated with $\varphi_n^{}$
\\
\cline{2-3}
&$\ff_n(z)$ & KK eigenfunction associated with $\psi_n^{}$
\\
\cline{2-3}
&$\ms;\mf$ & Bulk mass for scalar and fermion fields
\\
\cline{2-3}
&$M_{\s,n}^{};M_{\f,n}^{}$ & KK mass eigenvalue for scalar and fermion fields
\\
\cline{2-3}
&
$m_{\s,n}^{};m_{\f,n}^{}$
& KK mass for scalar and fermion fields
\\ \hline
\multirow{4}{*}{Other} 
&$\CC_k^{}$ & Color factors
\\
\cline{2-3}
&$\NN_k^{}$ & Kinemtic numerators
\\
\cline{2-3}
&$\C{a}{\cdots}$ & Wavefunction coupling coefficients
\\
\cline{2-3}
&$\tPhi$ & A set of all other possible physical states
\\ \hline
\end{tabular}
\caption{A reference table of notations used in the main text.}
\label{tab:1}
\end{table}

\newpage
\subsection{Kinematic Configuration}
\label{app:A.2}

The 4d gamma matrices $\ga^\mu$ in Dirac representation are given by
\begin{alignat}{3}
\label{eq:ga}
\ga^{0} \!=\! \begin{pmatrix}
1& 0& 0& 0\\
0& 1& 0& 0\\
0& 0& -1& 0\\
0& 0& 0& -1
\end{pmatrix}\!,~\,
\ga^{1}\!=\!\begin{pmatrix}
{0} & {0} & {0} & {1} \\ 
{0} & {0} & {1} & {0} \\ 
{0} & {-1} & {0} & {0} \\ 
{-1} & {0} & {0} & {0}
\end{pmatrix}\!,~\,
\ga^{2} \!=\! \begin{pmatrix}
{0} & {0} & {0} & {-\ii} \\ 
{0} & {0} & {\ii} & {0} \\ 
{0} & {\ii} & {0} & {0} \\ 
{-\ii} & {0} & {0} & {0}
\end{pmatrix}\!,~\,
\ga^{3}\!=\!\begin{pmatrix}
{0} & {0} & {1} & {0} \\ 
{0} & {0} & {0} & {-1} \\ 
{-1} & {0} & {0} & {0} \\ 
{0} & {1} & {0} & {0}
\end{pmatrix}\!,
\end{alignat}
and the fifth gamma matrix is
$\ga^5 = \ga_5^{} = \ii \ga^0\ga^1\ga^2\ga^3$.

\vs

In order to compute the amplitudes explicitly, we choose the momenta in the center-of-mass frame and make the initial state particles move along the $z$-axis.\ Then, the four-momenta for initial- and final-state particles are given by
\begin{equation}
\begin{alignedat}{3}
\label{eq:Momenta-2}
p_1^\mu &= -E(1, 0, 0, \be)  \,,  \qquad\qquad
& \hspace*{7mm}
&p_2^\mu = -E(1, 0, 0, -\be) \,,
\\[1mm]
p_3^\mu &= E ( 1, \be'\st, 0, \be'\ct) \,,
& \hspace*{7mm}
&p_4^\mu = E ( 1, -\be'\st, 0, -\be'\ct) \,,
\end{alignedat}
\end{equation}
where $\be=(1-m_{\s/\f,n}^2/E^2)^{\frac{1}{2}}\hs$ and 
$\be^\pp=(1-\Mmm/E^2)^{\frac{1}{2}}$.\footnote{For KK gravitons and gravitational Goldsone bosons, replacing $\Mm$ with $\MGm$.
}
With these, we can define the Mandelstam variables:
\begin{equation}
s =-( p_1^{} \!+\hsm p_2^{} )^2 \,, \qquad~
t =-( p_1^{} \!+\hsm p_4^{} )^2 \,, \qquad~
u =-( p_1^{} \!+\hsm p_3^{} )^2 \,,
\end{equation}
from which we have 
$\,s + t + u =2(m_{\s/\f,n}^2+\Mmm)$.
We further define the massless Mandelstam variables $(\sz,\tz,\uz)$ as:
\begin{equation}
\sz = 4E^2\be^2 \hs, \qquad~~
\tz = -\frac{\,\sz\,}{2}(1\!+\hsm\ct)\hs, \qquad~~
\uz  = -\frac{\,\sz\,}{2}(1\!-\hsm\ct)\hs,
\end{equation} 
where the sum of these Mandelstam variables is given by $\hs \sz+\tz+\uz=0\hs$.\ 
Using these, the spinors for initial-state KK fermions moving along the positive the positive $z$-axis $(\theta,\phi)=0$ or the negative $z$-axis $(\theta,\phi)=\pi$ are given by
\beqs 
\label{eq:SpinorInPsi0}
\begin{alignat}{3}
u^\uparrow_n  &=
(E\!+\!m_{\f,n}^{})^{\frac{1}{2}}
\begin{pmatrix}
{1} \\ {0} \\ \frac{E\be}{E+m_{\f,n}^{}} \\0
\end{pmatrix},
&\qquad& u^\downarrow_n
=(E\!+\!m_{\f,n}^{})^{\frac{1}{2}}
\begin{pmatrix}
{0} \\ {1} \\ {0} \\ -\frac{E\be}{E+m_{\f,n}^{}}
\end{pmatrix},  
\\[1mm]
v^\uparrow_n
&=(E\!+\!m_{\f,n}^{})^{\frac{1}{2}}
\begin{pmatrix}
\frac{E\be}{E+m_{\f,n}^{}} \\ 0 \\ -1 \\ 0
\end{pmatrix}  \,, \quad 
&&v^\downarrow_n
=(E\!+\!m_{\f,n}^{})^{\frac{1}{2}}
\begin{pmatrix}
0 \\ -\frac{E\be}{E+m_{\f,n}^{}} \\ 0 \\ -1
\end{pmatrix}\,,
\end{alignat}
\eeqs
where in the high energy limit, we denote $u_n^\pm = u_n^{\uparrow\downarrow}$ and $v_n^\pm=v_n^{\uparrow\downarrow}$.

\vs

Finally, the polarization vectors for final-state spin-1 particles are given by
\beqs
\begin{alignat}{3}
\ep^\mu_{3\hs\pm} &= \frac{1}{\sqrt{2}}(0, \pm\ct, -\ii, \mp\st)\,,
\hspace*{15mm}
&&\ep^\mu_{3\L} = \frac{E}{\Mm} (\be', \st, 0, \ct) \,,
\\
\ep^\mu_{4,\hs\pm} &=\frac{1}{\sqrt{2}}(0, \mp \ii\hs\ct, -\ii, \pm \st)\,,
\hspace*{8mm}
&&\ep^\mu_{4,\L} = \frac{E}{\Mm} (\be', -\st, 0, -\ct ) \,,
\end{alignat}
\eeqs
and the polarization tensors for final-state spin-2 particles are given by
\beqs
\begin{alignat}{3}
\vep_{3\pm1}^{\mn} & =
\frac{1}{\sqrt{2}}(\ep^{\mu}_{3\pm}\ep^{\nu}_{4\L} +
\ep^{\mu}_{3\pm}\ep^{\nu}_{3\L}) \,,\hspace*{12mm}&&
\vep_{3\L}^{\mn} =
\frac{1}{\sqrt{6}}(\ep^{\mu}_{3+}\ep^{\nu}_{3-}+\ep^{\mu}_{3-}\ep^{\nu}_{3+} + 2\ep^{\mu}_{3\L}\ep^{\nu}_{3\L}) \,,
\\
\vep_{4\pm1}^{\mn} & =
\frac{1}{\sqrt{2}}(\ep^{\mu}_{4\pm}\ep^{\nu}_{4\L} +
\ep^{\mu}_{4\pm}\ep^{\nu}_{4\L}) \,,&&
\vep_{4\L}^{\mn} =
\frac{1}{\sqrt{6}}(\ep^{\mu}_{4+}\ep^{\nu}_{4-}+\ep^{\mu}_{4-}\ep^{\nu}_{4+} + 2\ep^{\mu}_{4\L}\ep^{\nu}_{4\L}) \,,
\end{alignat}
\eeqs
where the eigenvalue of KK mass in each spin-1 longitudinal polarization vector $\ep^{\mu}_{i\L}$ should be replace by $\MM_n$.

\vs

In Section\,\ref{sec:2}, for the discussion of GAET, we have defined the following fields:
\begin{equation}
A_n^{a\L}=\ep_\mu^\L A_n^{a\mu}\,,\qquad~~~
v^a_n=v_\mu^{} A_n^{a\mu} , 
\end{equation}
where
\begin{equation}
\ep_\L^{\mu} = \ep_\S^\mu + v^\mu \,, \qquad
\ep_\S^\mu= p^\mu/\Mn\,, \qquad
v^{\mu}\!=\mO(1/E_n^{})\,.
\end{equation}
In GAET \eqref{eq:5D-ETI3-ALQ}, the residual term $\TT_v^{}$ is suppressed in the high-energy limit and can be treated as subleading, as it involves external gauge bosons contracted with the vector $v^\mu$.

\vs

In the discussion of GRET Type-I in Section\,\ref{sec:2}, we have defined the following gravitational fields:
\begin{equation}
h_n^{\pm1} \!= \vep^{\pm1}_{\mn} h^{\mn}_n , \qquad~~~
V_n^{\pm1} \!= \ep^{\pm}_{\mu} V^{\mu}_n
\hs, \qquad~~~
v_n^{\pm1} \!=\hsm v_{\mn}^{\pm1} h^{\mn}_n \hs.
\end{equation}
Under high energy expansion, we take $\ep_\L^{\mu}=\ep_\S^{\mu}+v^\mu$
with $\ep_{\S}^{\mu}\!=\!p^\mu\!/\MGn$ and $v^\mu\!=\hsm \mO(1/E_n^{})\hs$.\ 
With these, the polarization tensor $\vep_{\pm 1}^{\mn}$ can be decomposed as:
\begin{align}
\vep_{\pm 1}^{\mn} &= \vep_{\rm{S},\pm 1}^{\mn} + v_{\pm1}^{\mn}\,,\qquad~~
\vep_{\S,\pm 1}^{\mn} =\! \frac{1}{\sqrt{2}}
\big(\ep_\S^\mu\ep^\nu_{\pm} + \ep_\S^\nu\ep^\mu_{\pm}\big)\,,
\nn\\[1mm]
v_{\pm1}^{\mn} &=\! \frac{1}{\sqrt{2}}
\big(v^\mu\ep^\nu_{\pm} + v^\nu\ep^\mu_{\pm}\big)=\mO(1/E_n^{})\,. 
\end{align}
Therefore, the energy scaling of the residual term $\M_v^{}$ in \eqrefe{eq:KK-GRET-hV} is suppressed by the external states $v_n^{\pm1}$ and can be treated as subleading.

\vs

In the discussion of GRET Type-II, we have defined
\begin{equation}
h^\L_n = \vep^\L_{\mn}h^{\mn}_n \,,\qquad~~~
\tilde{\Delta}_n^{} = \frac{1}{\sqrt{6}}h_n^{} + \vt_n^{}\,,\qquad~~~
\vt_n^{}=\vt_{\mn}^{}h_n^{\mn}\,,
\end{equation}
where
\begin{align}
\vep_\L^{\mn} &=\frac{1}{\sqrt{6}}\!
\(\ep_+^\mu\ep_-^\nu \!+\ep_+^\nu\ep_-^\mu \!+ 2\ep_\L^\mu\ep_\L^\nu\)
= \sqrt{\frac{\,2\,}{3}}\,\vep_\S^{\mn} \!+ \vt^{\mn}\,,
\nn\\
\vt^{\mn} &= \vep_\L^{\mn}-\sqrt{\frac{\,2\,}{3}}\,\vep_\S^{\mn}
= \frac{1}{\sqrt{6}}\!\[\big(\ep_+^\mu\ep_-^\nu+ \ep_+^\nu\ep_-^\mu\big) 
+2\big(\ep_\S^\mu v^\nu +\ep_\S^\nu v^\mu + v^\mu v^\nu\big)\]=\mO(E_n^0)\,.
\end{align}
As demostrated in Ref.\,\cite{Hang:2024uny}, the external state $\tilde{\Delta}_n^{}$ can be further expressed as:
\begin{equation}
\tilde{\Delta}_n^{}= \sqrt{\frac{\,3\,}{2}}\Big(2\hs\bC_{\rm{mod}}^{V,nm}\hs v_{\mu}^{}V^{\mu}_m
+ v_{\mu}^{}v_{\nu}^{}h_n^{\mn}\Big)\,, 
\end{equation}
where the two terms on right-hand side are suppressed by the factors of 
$v^{\mu}\!=\!\mO(1/E_n^{})$ and $v^{\mu}v^{\nu}\!=\!\mO(1/E^{2}_n)$, individually.\ Consequently, the residual term $\M_\Delta^{}$ in \eqrefe{eq:KK-GRET-hphi} is suppressed by those external-state factors and can be regarded as subleading.

\section{Wavefunctions}
\label{app:B}

\subsection{KK Gauge and Gravity Wavefunction Solutions}
\label{app:B.1}

\paragraph{Gauge Sector}
The wavefunctions $\gg_n^{}(z)$ and $\tgg_n^{}(z)$ can be solved 
from the equations of motion \eqref{eq:EOM-gg-tgg}
in terms of Bessel functions:
\beqs
\label{Aeq:gn-gtn}
\begin{align}
\label{Aeq:gn}	
\gg_n^{}(z) &= \frac{e^{-\A(z)}}{N_n^{\rm{ga}}}
\Big[J_1^{}\Big(e^{-\A(z)}\Mn/k\Big) + 
b_{n0}^{}\hs Y_1^{}\Big(e^{-\A(z)}\Mn/k\Big)\Big]\,,
\\[1mm]
\label{Aeq:gtn}	
\tgg_n^{}(z) &= \frac{e^{-\A(z)}}{{N}_n^{\rm{ga}}}
\Big[J_0^{}\Big(e^{-\A(z)}\Mn/k\Big) + 
b_{n0}^{}\hs Y_0^{}\Big(e^{-\A(z)}\Mn/k\Big)\Big]\,,
\end{align}
\eeqs
where the normalization factors $N_n^{\rm{ga}}$ 
can be fixed by the orthonormal conditions \eqref{eq:Normalize-YM}.\
The coefficient $b_{n0}^{}$ is derived as follows:
\begin{equation}
\label{Aeq:bn0}	
b_{n0}^{}=- \frac{J_0^{}\big(\Mn/k\big)}{Y_0^{}\big(\Mn/k\big)} \,,
\end{equation}
The KK mass $\Mn$ is determined by the roots of the following equation:
\begin{equation}
\label{Aeq:YM-Mn}
J_0^{}\Big(e^{-\A(L)}\Mn/k\Big)+b_{n0}^{} Y_0^{}\Big(e^{-\A(L)}\Mn/k\Big) = 0 \,.
\end{equation}
For the massless zero-mode of wavefunction $\gg_0^{}$, it can be solved from 
\eqrefe{eq:EOM-gg-tgg}:
\begin{equation}
\label{Aeq:g0}	
\gg_0^{} = \Big[\Big(1-e^{-\A(L)}\Big)/\A(L)\,\Big]^{\frac{1}{2}}\,.
\end{equation}

\paragraph{Gravity Sector}
Solving the equations of motion \eqref{eq:EOM-un-vn} and \eqref{eq:EOM-wn}, we are able to express the wavefunctions $\uu_n^{}(z),\hs\vv_n^{}(z)$ and $\ww_n^{}(z)$
in terms of the Bessel functions:
\beqs
\label{Aeq:KK-uvw} 
\begin{align}
\label{Aeq:KK-un} 
\uu_n^{}(z) &=
\frac{e^{-2\A(z)}}{N_n^{\rm{gr}}}\Big[J_2^{}\Big(e^{-\A(z)}\MGn/k\Big)+b_{n1}^{} Y_2^{}\Big(e^{-\A(z)}\MGn/k\Big)\Big] \,,
\\[1mm]
\label{Aeq:KK-vn} 
\vv_n^{}(z) &=
\frac{e^{-2\A(z)}}{N_n^{\rm{gr}}}\Big[J_1^{}\Big(e^{-\A(z)}\MGn/k\Big)+b_{n1}^{} Y_1^{}\Big(e^{-\A(z)}\MGn/k\Big)\Big]\,,
\\[1mm]
\label{Aeq:KK-wn} 
\ww_n^{}(z) &=
\frac{e^{-2\A(z)}}{N_n^{\rm{gr}}}\Big[J_0^{}\Big(e^{-\A(z)}\MGn/k\Big)+b_{n1}^{}Y_0^{}\Big(e^{-\A(z)}\MGn/k\Big)\Big]\,,
\end{align}
\eeqs
where $N_n^{\rm{gr}}$ is the normalization factor fixed by orthonormal conditions given in Eqs.\,\eqref{eq:Normalize-u-v} and \eqref{eq:EOM-normalize-w}.\ Further, the coefficient $b_{n1}^{}$ is given by
\begin{equation}
b_{n1}^{}= -\frac{J_1^{}\big(\MGn/k\big)}{Y_1^{}\big(\MGn/k\big)}\,.
\end{equation}
The KK mass eigenvalue $\MGn$ is determined by solving the equation:
\begin{equation}
\label{Aeq:GR-Mn}
J_1^{}\Big(e^{-\A(L)}\MGn/k\Big) +b_{n1}^{}Y_1^{}\Big(e^{-\A(L)}\MGn/k\Big) = 0 \,.
\end{equation}
The zero-mode wavefunctions $\uu_0^{}(z)$ and $\ww_0^{}(z)$
can be solved as follows:
\beqs
\label{Aeq:u0-w0}
\begin{align}
\label{Aeq:u0}
\uu_0^{} &=\sqrt{2}\,\Big[\,e^{\A(L)}\!+e^{2\A(L)}\,\Big]^{-\frac{1}{2}}\,,
\\[0mm]
\label{Aeq:w0}
\ww_0^{}(z) &=\sqrt{2}\,e^{-2\A(z)}\Big[\,1+e^{-\A(L)}\,\Big]^{-\frac{1}{2}}\,.
\end{align}
\eeqs

\paragraph{Scalar Sector} 
Solving \eqrefe{eq:S-EOM}, we can obtain the following equations:
\beqs
\label{Aeq:s0-sn}
\begin{align} 
\ss_0^{}(z) &= \big(1+4k^2/\ms^2\,\big)e^{-2\A(z)} \,,
\\[1mm]
\ss_n^{}(z) &=
\frac{e^{-2\A(z)}}{N_n^\s}\Big[J_\al^{}\Big(e^{-\A(z)}\Msn/k\Big)+b_{n\al}^{} Y_\al^{}\Big(e^{-\A(z)}\Msn/k\Big)\Big]\,,
\end{align} 
\eeqs
where $N_n^{\s}$ is the normalization factor and $b_{n\al}^{}$ is given by
\begin{equation}
b_{n\al}^{}=-\frac{2J_\al^{}\big(\Msn/k\big)+\big(\Msn/k\big)J^\pp_\al\big(\Msn/k\big)}{2Y_\al^{}\big(\Msn/k\big)+\big(\Msn/k\big)Y^\pp_\al\big(\Msn/k\big)}\,, \qquad~~
\al = \(4+\frac{\ms^2}{k^2}\)^{\!\!\frac{1}{2}}.
\end{equation}
The KK mass $\Msn$ is determined by roots of the eigenvalue equation:
\begin{align}
\label{Aeq:S-Msn}
&\xoverline{M}_{\s,n}^{} J^\pp_\al(\xoverline{M}_{\s,n}^{})\Big[2Y_\al(\widetilde{M}_{\s,n}^{})\!+\!\widetilde{M}_{\s,n}^{}Y^\pp_\al(\widetilde{M}_{\s,n}^{})\Big]\!+2J_\al(\xoverline{M}_{\s,n}^{})\Big[2Y_\al(\widetilde{M}_{\s,n}^{})\!+\! \widetilde{M}_{\s,n}^{} Y^\pp_\al(\widetilde{M}_{\s,n}^{})\Big]
\nn\\
&-(\xoverline{M}_{\s,n}^{}\leftrightarrow\widetilde{M}_{\s,n}^{})=0\,,
\end{align}
where $\xoverline{M}_{\s,n}^{}=\Msn/k$ and $\widetilde{M}_{\s,n}^{}=e^{-\A(L)}\Msn/k$\,.

\paragraph{Fermion Sector} 
Solving the equations of motion \eqref{eq:EOM-dn-kn}, we have
\beqs
\label{Aeq:d0-dn-kn}
\begin{align} 
\dd_0^{}(z) &= (1+kz)^{\frac{\mf}{k}}e^{-\mf z}e^{-2\A(z)} \,,
\\[1mm]
\dd_n^{}(z) &=
\frac{e^{-5\A(z)/2}}{N_n^\f}\Big[J_{\al_-}^{}\Big(e^{-\A(z)}\Mfn/k\Big)+b_{n\al_-}Y_{\al_-}^{}\Big(e^{-\A(z)}\Mfn/k\Big)\Big]\,,
\\[1mm]
\kk_n^{}(z) &=
\frac{e^{-5\A(z)/2}}{N_n^\f}\Big[J_{\al_+}^{}\Big(e^{-\A(z)}\Mfn/k\Big)+b_{n\al_+} Y_{\al_+}^{}\Big(e^{-\A(z)}\Mfn/k\Big)\Big]\,,
\end{align} 
\eeqs
where $N_n^{\f}$ is the normalization factor and $b_{n\al_\pm}$ are given by
\beqs
\begin{align}
b_{n\al_+}&=-\frac{J^{}_{\al_+}\big(\Mfn/k\big)}
{Y^{}_{\al_+}\big(\Mfn/k\big)}\,, \qquad~~
\al_+ = \frac{1}{2}+\frac{\mf^{}}{k}\,,
\\[1mm]
b_{n\al_-}&=-\frac{\al_-J_{\al_-}^{}\big(\Mfn/k\big)+\big(\Mfn/k\big)J^\pp_{\al_-}\big(\Mfn/k\big)}
{\al_-Y_{\al_-}^{}\big(\Msn/k\big)+\big(\Mfn/k\big)Y^\pp_{\al_-}\big(\Mfn/k\big)}\,, \qquad~~
\al_- = \frac{1}{2}-\frac{\mf^{}}{k}\,,
\end{align}
\eeqs
and the KK mass $\Mfn$ is dertermined by
\begin{equation}
\label{Aeq:F-Mfn}
J_{\al_+}^{}\Big(e^{-\A(L)}\Mfn/k\Big)+b_{n\al_+} Y_{\al_+}^{}\Big(e^{-\A(L)}\Mfn/k\Big) = 0\,.
\end{equation}

\subsection{Wavefunction Coupling Coefficients and Sum Rule Identities}
\label{app:B.2}

The general wavefunction coupling coefficient is defined as:
\begin{equation}
\label{eq:C-a}
\C{a}{\XX_n^{}\YY_m^{}\mathbf{Z}_\ell^{}\cdots}= \frac{1}{L}\int_0^L\!\td z\,e^{a\A(z)}\, \XX_n^{}(z)\hs\YY_m^{}(z)\hs\mathbf{Z}_\ell^{}(z)\cdots\,,
\end{equation}
where $\XX_n^{},\YY_m^{},\mathbf{Z}_\ell^{},\cdots$ denote the eigenfunctions corresponding to the KK fields, including possible zero-mode fields.\
This integral integrates over the 5d coordinate $z$, giving a numerical coefficient once the background geometry, compactification size and the KK masses are specified.\

\vs

As an example, consider the 4d three-KK gauge boson vertex
\begin{equation}
\VV[A^{a\mu}_{n_1}A^{b\nu}_{n_2}A^{c\rho}_{n_3}] = - g\hs \C{1}{\gg_{n_1}^{}\gg_{n_2}^{}\gg_{n_3}^{}} f^{abc}\big[\, \eta^{\mn}(p_1^{}\!-p_2^{})^\rho+\rm{cyc.}\,\big] \,,
\end{equation}
where each eigenfunction $\gg_{n_i}^{}$ is expressed in terms of Bessel form given in \eqrefe{Aeq:gn}, with the eigenmass $M_{n_i}^{}$ determined by the root of \eqrefe{Aeq:YM-Mn}.\
By fixing the geometry (choosing the scale $\bk$, the length $L$, and solving for $M_{n_i}^{}$), the integral of $\C{1}{\gg_{n_1}^{}\gg_{n_2}^{}\gg_{n_3}^{}}$ will result in an explicit numerical value.\
Moreover, in the flat-space limit $\bk\ito0$, all the functions $\gg_{n_i}^{}$ reduce to simple trigonometric forms \cite{Hang:2021fmp,Hang:2022rjp}:
\begin{equation}
\gg_{n_i}^{}=\sqrt{2}\cos\big(M_{n_i}^{} z\big)\,,\qquad
\rm{with}~~M_{n_i}^{}=\frac{n_i\hs\pi}{L}\,.
\end{equation}
In this case, the wavefunction coupling $\C{1}{\gg_{n_1}^{}\gg_{n_2}^{}\gg_{n_3}^{}}$ becomes
\begin{equation}
\label{Aeq:C-1-nml}
\C{1}{\gg_{n_1}^{}\gg_{n_2}^{}\gg_{n_3}^{}} = \frac{1}{\sqrt{2}}\big[\delta({n_1}\!+\!{n_2}\!-\!{n_3})+\delta({n_1}\!-\!{n_2}\!+\!{n_3})+\delta({n_1}\!-\!{n_2}\!-\!{n_3}) \big] \,,
\end{equation}
which indicates a selection rule for KK modes.\
Now, if we take $(n_1,n_2,n_3)=(1,1,2)$, only the first delta function in \eqrefe{Aeq:C-1-nml} survives, yielding
\begin{equation}
\C{1}{\gg_1^{}\gg_1^{}\gg_2^{}} = \frac{1}{\sqrt{2}}\,,
\end{equation}
where it shows that once the mode profiles and their masses are fixed, the wavefunction coupling is just an explicit calculable number.

\vs

The wavefunction coefficients also obey a set of the sum rule relations, which we have used in Section\,\ref{sec:3}. 
For completeness, we provide explicit derivation of these sum rules.\
To derive the sum rule \eqref{eq:SR-s-1}, we start from the three-point GAET indentity \eqref{eq:GAET-3pt}, choosing $(n_1,n_2,n_3)=(m,m,\jj)$, which gives
\begin{equation}
\label{Aeq:GAET-3pt}
\big(2\Mmm-\Mjj\big) \C{1}{\gg_{m}^2\gg_{\jj}^{}}
= 2 \hs \Mmm \C{1}{\tgg_{m}^{2}\gg_{\jj}^{}}\,.
\end{equation}
Multiplying both side by $\C{3}{\ss_n^2\hs\gg_\jj^{}}$ and sum over the KK index $\jj$, we obtain
\begin{equation}
\sum_{\jj=0}^{\infty}M_\jj^2\hs \C{3}{\ss_n^2\hs\gg_\jj^{}}\hs \C{1}{\gg_m^2\gg_\jj^{}} = 2M_m^2 \Big(\C{3}{\ss_n^2\hs\gg_m^2}-\C{3}{\ss_n^2\hs\tgg_m^2}\Big)\,,
\end{equation}
where the completeness relation \eqref{eq:completness-1} is used on the right-hand side to combine the product of two wavefunction coefficients into a single term.

\vs

For the sum rule \eqref{eq:SR-s-2}, we can derive
\begin{align}
\label{eq:SR-s-2-tem1}
&\quad\,\sum_{\jj=0}^{\infty} M_{\s,\jj}^2 \C{3}{\ss_n^{}\gg_m^{}\ss_\jj^{}}\C{3}{\ss_n^{}\gg_m^{}\ss_\jj^{}}
\nn\\
&=\sum_{\jj=0}^{\infty} \C{3}{\ss_n^{}\gg_m^{}\ss_\jj^{}} \C{3}{\ss_n^{} \gg_m^{} (M_{\s,\jj}^2\ss_\jj^{})}
\nn\\
&= \sum_{\jj=0}^{\infty} \C{3}{\ss_n^{}\gg_m^{}\ss_\jj^{}} \Big(\C{3}{\pd_z^{}\ss_n^{}\gg_m^{}\pd_z^{}\ss_\jj^{}}
+\C{3}{\ss_n^{}\pd_z^{}\gg_m^{}\pd_z^{}\ss_\jj^{}}
+m_\s^2\C{5}{\ss_n^{}\gg_m^{}\ss_\jj^{}}\Big)
\nn\\
&= \Mmm\C{3}{\ss_n^{2}\gg_m^{2}}-
2\C{3}{\ss_n^{}\pd_z^{}\ss_n^{}\gg_m^{}\pd_z^{}\gg_m^{}}
+M_{\ss,n}^2\C{3}{\ss_n^{2}\gg_m^{2}}
-2\C{3}{\A^\pp\ss_n^{2}\gg_m^{}\pd_z^{}\gg_m^{}}\,,
\end{align}
where the EOMs for gauge and scalar fields, \eqrefe{eq:EOM-gg-tgg} and \eqrefe{eq:S-EOM}, together with integration by parts, have been applied to obtain the third line of \eqrefe{eq:SR-s-2-tem1}, followed by the use of the completeness relation \eqref{eq:completness-1} to arrive at the last line.\ 
Further, for the first term in the last line of \eqrefe{eq:SR-s-2-tem1}, we invoke the \eqrefe{eq:EOM-gg-tgg} once more to derive
\begin{align}
\label{eq:SR-s-2-tem2}
\C{3}{\ss_n^{2}(\Mmm\gg_m^{})\gg_m^{}}
&= \C{3}{\ss_n^2 (\pd_z^{}\gg_m^{})^2\,} 
+2\C{3}{\ss_n^{}\pd_z^{}\ss_n^{}\gg_m^{}\pd_z^{}\gg_m^{}}
+2\C{3}{\A^\pp\ss_n^{2}\gg_m^{}\pd_z^{}\gg_m^{}}
\nn\\[1mm]
&=\Mmm\C{3}{\ss_n^2\tgg_m^2} 
+2\C{3}{\ss_n^{}\pd_z^{}\ss_n^{}\gg_m^{}\pd_z^{}\gg_m^{}}
+2\C{3}{\A^\pp\ss_n^{2}\gg_m^{}\pd_z^{}\gg_m^{}}\,,
\end{align}
where \eqrefe{eq:EOM-gg-tgg-2} has been used in the last step.\
Finally, substituting \eqrefe{eq:SR-s-2-tem2} into \eqrefe{eq:SR-s-2-tem1}, we can obtain the following identity
\begin{equation}
\sum_{\jj=0}^{\infty} M_{\s,\jj}^2 \Big(\C{3}{\ss_n^{}\gg_m^{}\ss_\jj^{}}\Big)^2  
= \Mmm \C{3}{\ss_n^2\tgg_m^2} + M_{\s,n}^2 \C{3}{\ss_n^2\gg_m^2}\,.
\end{equation}
The proofs of the sum rules \eqref{eq:SR-f-1}-\eqref{eq:SR-f-2} follow analogously by applying the corresponding EOMs, completeness relations and integration-by-parts method.

\section{Lagrangians}
\label{app:C}

In this appendix, we present the relevant interaction Lagrangians up to four-point vertices, categorized according to their field content and interaction type.

\paragraph{KK Scalar Interactions}
Three-point interaction Lagrangians involving KK scalars and gauge/gravity fields are expressed as follows:
\beqs
\begin{align}
\La[\varphi\varphi A_\mu] &= \ii g \sum_{n,m,\ell}
\big(\varphi^\ast_{i,n}\pd_\mu^{}\varphi^{}_{j,m}-\pd_\mu^{}\varphi^\ast_{i,n}\varphi^{}_{j,m}\big)A_{\ell}^{a\mu}T^a_{ij}\,\C{3}{\ss_n^{}\ss_m^{}\gg_\ell^{}}\,,
\\[1mm]
\La[\varphi\varphi A_5] &= \ii g\sum_{n,m,\ell}
\varphi^\ast_{i,n}\varphi^{}_{j,m}A_\ell^{a5}T^a_{ij}
\big(\C{3}{\ss_n^{}\ss_m'\tgg_\ell^{}}-\C{3}{\ss_n^{\pp}\ss_m^{}\tgg_\ell^{}}\big)\,,
\\[1mm]
\La[\varphi\varphi h] &=
-\frac{\ka}{2}\sum_{n,m,\ell}\big( 2\pd_\mu^{}\varphi_n^\ast\pd_\nu^{}\varphi_m^{}h^{\mn}_\ell
- \pd^\al\varphi_n^\ast\pd_\al^{}\varphi_m^{} h^{}_\ell\big)\C{3}{\ss_n^{}\ss_m^{}\uu_\ell^{}}\,.
\end{align}
\eeqs
The four-point interaction Lagrangians take the form:
\beqs
\begin{align}
\La[\varphi\varphi A_5A_5] &= g^2\sum_{n,m,\ell,q}
\varphi^\ast_{i,n}\varphi^{}_{j,m}A_\ell^{a5}A_q^{b5}T^a_{ik}T^b_{kj}\,
\C{3}{\ss_n^{}\ss_m^{}\tgg_\ell^{}\tgg_q^{}}\,,
\\[1mm]
\La[\varphi\varphi VV] &= -\frac{\ka^2}{4}\sum_{n,m,\ell,q}
\pd^\mu\varphi^\ast_{i,n}\pd_\mu^{}\varphi^{}_{j,m}V_\ell^{\al}V_{\al,q}^{}\,
\C{3}{\ss_n^{}\ss_m^{}\vv_\ell^{}\vv_q^{}} \,,
\\[1mm]
\La[\varphi\varphi\phi\phi] &= -\frac{\ka^2}{4}\sum_{n,m,\ell,q}
\pd^\mu\varphi^\ast_{i,n}\pd_\mu^{}\varphi^{}_{j,m}\phi_\ell^{}\phi_q^{}\,
\C{3}{\ss_n^{}\ss_m^{}\ww_\ell^{}\ww_q^{}} \,.
\end{align}
\eeqs

\paragraph{KK Fermion Interactions}
Three-point interaction Lagrangians involving KK fermions and gauge/gravity fields are given by
\beqs
\label{Aeq:La-f-3}
\begin{align}
&\La[\psi\psi A_\mu]= g\sum_n \Big(\,\bpsi^{(1)}_{i,n} \ga_\mu \psi_{j,0}^{}c^n_1
- \bpsi^{(2)}_{i,n} \ga_\mu \psi_{j,0}^{} d^{\hs n}_1
+h.c.\Big)A_n^{\mu a}T^a_{ij} \C{4}{\dd_0^{}\dd_n^{}\gg_n^{}}\,,
\\
&\La[\psi\psi A_5]=-\ii g \sum_{n}\Big(\bpsi^{(1)}_{i,n}  \psi_{j,0}^{} c_2^n
+ \bpsi^{(2)}_{i,n} \psi_{j,0}^{} d_2^{\hs n}+h.c.\Big)A^{5a}_n T^a_{ij} \,\C{4}{\dd_0^{}\kk_n^{}\tgg_n^{}}\,,
\\
&\La[\psi\psi h] =- \frac{\ii \ka}{4} \sum_{n}\Big[
h_n^{\mn} \Big(\bpsiI_n\ga_\mu c_1^{n}\pd_\nu\psi_0^{} - \pd_\nu \bpsiI_n\ga_\mu c_1^{n}\psi_0^{}-\bpsiII_n\ga_\mu d_1^{\hs n} \pd_\nu\psi_0^{} + \pd_\nu \bpsiII_n\ga_\mu d_1^{\hs n} \psi_0^{}\Big)
\nn\\[-.5mm]
&-h_n^{}\Big(\bpsiI_n\ga^{\mu} c_1^{n} \pd_\mu \psi_0^{} \!-\! \pd_\mu \bpsiI_n\ga^\mu c_1^{n} \psi_0^{} \!-\! \bpsiII_n\ga^\mu d_1^{\hs n} \pd_\mu \psi_0^{} \!+\! \pd_\mu \bpsiII_n\ga^\mu d_1^{\hs n} \psi_0^{}\Big) \!+ h.c.\Big] \C{4}{\dd_0^{}\dd_n^{}\uu_n^{}}
\nn\\[.5mm]
&-\frac{\ka}{4} \sum_n h_n^{} \Big(\bpsiI_n \psi_0^{}\hs c_2^{n} - \bar{\psi}_0^{} \psiI_n\hs d_2^{\hs n}+ \bpsiII_n \psi_0^{}\hs d_2^{\hs n} - \bar{\psi}_0^{} \psiII_n\hs c_2^{n} \Big)\C{4}{\dd_0^{}\kk_n^{\pp}\uu_n^{}} \,,
\\[1mm]
&\La[\psi\psi V]=
\frac{\ii\sqrt{2}\hs\ka}{8}\sum_n V_n^\mu \Big(\bpsiI_n c_2^n \pd_\mu \psi_0^{} - \pd_\mu \bpsiI_n c_2^n \psi_0^{} 
-\bpsiII_n d_2^{\hs n} \pd_\mu \psi_0^{} 
\nn\\[-1mm]
&\hspace*{3cm}+ \pd_\mu \bpsiII_n d_2^{\hs n} \psi_0^{} 
+ h.c.\Big) \C{4}{\dd_0^{}\dd_n^{}\vv_n^{}} \,,
\\[1mm]
&\La[\psi\psi\phi]= -\frac{\ii\sqrt{6}\hs\ka}{24}\sum_n\phin \Big(\bpsiI_n \ga^{\mu} c_1^n \pd_\mu \psi_0^{} - \pd_\mu \bpsiI_n\ga^\mu c_1^n \psi_0^{} 
-\bpsiII_n\ga^\mu d_1^{\hs n} \pd_\mu \psi_0^{} 
\nn\\[-1mm]
&\hspace*{3cm}+\pd_\mu \bpsiII_n\ga^\mu d_1^{\hs n} \psi_0^{}+ h.c.\Big) \C{4}{\dd_0^{}\dd_n^{}\ww_n^{}}\,,
\end{align}
\eeqs
where the parameters $(c_1^{n},c_2^{n},d_1^{\hs n},d_2^{\hs n})$ are defined as follows:
\begin{alignat}{3}
c_1^{n} &= (\cos\vtn P_L^{} + \sin\vtn P_{R}^{})  \,, \qquad~~~
&& d_1^{\hs n} = (\sin\vtn P_{L}^{} + \cos\vtn P_{R}^{})   \,,
\nn\\[1mm]
c_2^{n} &= (\cos\vtn P_{L}^{} - \sin\vtn P_{R}^{})  \,, \qquad~~~
&& d_2^{\hs n}= (\sin\vtn P_{L}^{} - \cos\vtn P_{R}^{})  \,,
\end{alignat}
with the projection operators and angle $\vtn$
\begin{equation}
P_{L/R}^{}= (1\pm\ga_5^{})/2\,, \qquad~~
\vtn = \frac{1}{2}\arctan\!\(\frac{e^{\A}\hs\mf^{}}{\Mfn}\).
\end{equation} 
The four-point interaction Lagrangians are given by
\beqs
\label{Aeq:La-f-4}
\begin{align}
\La[\psi\psi VV] &= -\frac{\ii\ka^2}{64}\sum_n\Big[16(V_n^\mu)^2 (\bpsi_0^{} \overset{\leftrightarrow{}}{\slpd}\psi_0^{})
-(V^\lam_n \pd^\mu V^\nu_n\!-\!V^\nu_n \pd^\mu V^\lam_n)
(\bpsi_0^{}\ga_\mu^{} \ga_\nu^{}\ga_\lam^{}\psi_0^{})\Big]
\nn\\
&\hspace*{1cm}\times\C{4}{\dd_0^2\vv_n^2} \,,
\\[1mm]
\La[\psi\psi\phi\phi] &= -\frac{\ii\ka^2}{4}\sum_n\phi_n^2(\bpsi_0^{} \ga^\mu\overset{\leftrightarrow{}}{\pd}_{\!\mu}\psi_0^{})\C{4}{\dd_0^2\ww_n^2}\,.
\end{align}
\eeqs
In \eqrefe{Aeq:La-f-3}, we focus primarily on interaction terms between one zero-mode and one $n$-mode ($n\!>\!0$) fermion.\ 
For the four-point interaction Lagrangians \eqref{Aeq:La-f-4}, we restrict our presentation to the terms involving two zero-mode fermions.\ 
For the interaction Lagrangians involving two non-zero KK mode fermions are analogous to those shown above, their expressions are omitted for brevity.

\newpage

\bibliographystyle{utphys}
\bibliography{RSET-Refs.bib}

\providecommand{\href}[2]{#2}\begingroup\raggedright\begin{thebibliography}{10}

\bibitem{Kaluza:1921tu}
T.~Kaluza, ``{Zum Unit\"atsproblem der Physik},''
  \href{http://dx.doi.org/10.1142/S0218271818700017}{{\em Sitzungsber. Preuss.
  Akad. Wiss. Berlin (Math. Phys. )} {\bfseries 1921} (1921) 966--972},
  \href{http://arxiv.org/abs/1803.08616}{{\ttfamily arXiv:1803.08616
  [physics.hist-ph]}}.

\bibitem{Klein:1926tv}
O.~Klein, ``{Quantum Theory and Five-Dimensional Theory of Relativity. (In
  German and English)},'' \href{http://dx.doi.org/10.1007/BF01397481}{{\em Z.
  Phys.} {\bfseries 37} (1926) 895--906}.

\bibitem{Chivukula:2001esy}
R.~S. Chivukula, D.~A. Dicus, and H.-J. He, ``{Unitarity of compactified
  five-dimensional Yang-Mills theory},''
  \href{http://dx.doi.org/10.1016/S0370-2693(01)01435-6}{{\em Phys. Lett. B}
  {\bfseries 525} (2002) 175--182},
  \href{http://arxiv.org/abs/hep-ph/0111016}{{\ttfamily arXiv:hep-ph/0111016}}.

\bibitem{Chivukula:2002ej}
R.~S. Chivukula and H.-J. He, ``{Unitarity of deconstructed five-dimensional
  Yang-Mills theory},''
  \href{http://dx.doi.org/10.1016/S0370-2693(02)01495-8}{{\em Phys. Lett. B}
  {\bfseries 532} (2002) 121--128},
  \href{http://arxiv.org/abs/hep-ph/0201164}{{\ttfamily arXiv:hep-ph/0201164}}.

\bibitem{He:2004zr}
H.-J. He, ``{Higgsless deconstruction without boundary condition},''
  \href{http://dx.doi.org/10.1142/S0217751X05026583}{{\em Int. J. Mod. Phys. A}
  {\bfseries 20} (2005) 3362--3380},
  \href{http://arxiv.org/abs/hep-ph/0412113}{{\ttfamily arXiv:hep-ph/0412113}}.

\bibitem{Hang:2021fmp}
Y.-F. Hang and H.-J. He, ``{Structure of Kaluza-Klein graviton scattering
  amplitudes from the gravitational equivalence theorem and double copy},''
  \href{http://dx.doi.org/10.1103/PhysRevD.105.084005}{{\em Phys. Rev. D}
  {\bfseries 105} no.~8, (2022) 084005},
  \href{http://arxiv.org/abs/2106.04568}{{\ttfamily arXiv:2106.04568
  [hep-th]}}.

\bibitem{Hang:2022rjp}
Y.-F. Hang and H.-J. He, ``{Gravitational Equivalence Theorem and Double-Copy
  for Kaluza-Klein Graviton Scattering Amplitudes},''
  \href{http://dx.doi.org/10.34133/2022/9860945}{{\em Research} {\bfseries
  2022} (2022) 9860945}, \href{http://arxiv.org/abs/2207.11214}{{\ttfamily
  arXiv:2207.11214 [hep-th]}}.

\bibitem{Hang:2024uny}
Y.~Hang, W.-W. Zhao, H.-J. He, and Y.-L. Qiu, ``{Structure of massive
  gauge/gravity scattering amplitudes, equivalence theorems, and extended
  double-copy with compactified warped space},''
  \href{http://dx.doi.org/10.1007/JHEP02(2025)001}{{\em JHEP} {\bfseries 02}
  (2025) 001}, \href{http://arxiv.org/abs/2406.12713}{{\ttfamily
  arXiv:2406.12713 [hep-th]}}.

\bibitem{Chivukula:2023qrt}
R.~S. Chivukula, J.~A. Gill, K.~A. Mohan, D.~Sengupta, E.~H. Simmons, and
  X.~Wang, ``{Symmetries, spin-2 scattering amplitudes, and equivalence
  theorems in warped five-dimensional gravitational theories},''
  \href{http://dx.doi.org/10.1103/PhysRevD.109.075016}{{\em Phys. Rev. D}
  {\bfseries 109} no.~7, (2024) 075016},
  \href{http://arxiv.org/abs/2312.08576}{{\ttfamily arXiv:2312.08576
  [hep-ph]}}.

\bibitem{Randall:1999ee}
L.~Randall and R.~Sundrum, ``{A Large mass hierarchy from a small extra
  dimension},'' \href{http://dx.doi.org/10.1103/PhysRevLett.83.3370}{{\em Phys.
  Rev. Lett.} {\bfseries 83} (1999) 3370--3373},
  \href{http://arxiv.org/abs/hep-ph/9905221}{{\ttfamily arXiv:hep-ph/9905221}}.

\bibitem{Randall:1999vf}
L.~Randall and R.~Sundrum, ``{An Alternative to compactification},''
  \href{http://dx.doi.org/10.1103/PhysRevLett.83.4690}{{\em Phys. Rev. Lett.}
  {\bfseries 83} (1999) 4690--4693},
  \href{http://arxiv.org/abs/hep-th/9906064}{{\ttfamily arXiv:hep-th/9906064}}.

\bibitem{Kawai:1985xq}
H.~Kawai, D.~C. Lewellen, and S.~H.~H. Tye, ``{A Relation Between Tree
  Amplitudes of Closed and Open Strings},''
  \href{http://dx.doi.org/10.1016/0550-3213(86)90362-7}{{\em Nucl. Phys. B}
  {\bfseries 269} (1986) 1--23}.

\bibitem{Bern:2008qj}
Z.~Bern, J.~J.~M. Carrasco, and H.~Johansson, ``{New Relations for Gauge-Theory
  Amplitudes},'' \href{http://dx.doi.org/10.1103/PhysRevD.78.085011}{{\em Phys.
  Rev. D} {\bfseries 78} (2008) 085011},
  \href{http://arxiv.org/abs/0805.3993}{{\ttfamily arXiv:0805.3993 [hep-ph]}}.

\bibitem{Bern:2010ue}
Z.~Bern, J.~J.~M. Carrasco, and H.~Johansson, ``{Perturbative Quantum Gravity
  as a Double Copy of Gauge Theory},''
  \href{http://dx.doi.org/10.1103/PhysRevLett.105.061602}{{\em Phys. Rev.
  Lett.} {\bfseries 105} (2010) 061602},
  \href{http://arxiv.org/abs/1004.0476}{{\ttfamily arXiv:1004.0476 [hep-th]}}.

\bibitem{Bern:2019prr}
Z.~Bern, J.~J. Carrasco, M.~Chiodaroli, H.~Johansson, and R.~Roiban, ``{The
  duality between color and kinematics and its applications},''
  \href{http://dx.doi.org/10.1088/1751-8121/ad5fd0}{{\em J. Phys. A} {\bfseries
  57} no.~33, (2024) 333002}, \href{http://arxiv.org/abs/1909.01358}{{\ttfamily
  arXiv:1909.01358 [hep-th]}}.

\bibitem{Johansson:2014zca}
H.~Johansson and A.~Ochirov, ``{Pure Gravities via Color-Kinematics Duality for
  Fundamental Matter},'' \href{http://dx.doi.org/10.1007/JHEP11(2015)046}{{\em
  JHEP} {\bfseries 11} (2015) 046},
  \href{http://arxiv.org/abs/1407.4772}{{\ttfamily arXiv:1407.4772 [hep-th]}}.

\bibitem{Johansson:2015oia}
H.~Johansson and A.~Ochirov, ``{Color-Kinematics Duality for QCD Amplitudes},''
  \href{http://dx.doi.org/10.1007/JHEP01(2016)170}{{\em JHEP} {\bfseries 01}
  (2016) 170}, \href{http://arxiv.org/abs/1507.00332}{{\ttfamily
  arXiv:1507.00332 [hep-ph]}}.

\bibitem{Johansson:2019dnu}
H.~Johansson and A.~Ochirov, ``{Double copy for massive quantum particles with
  spin},'' \href{http://dx.doi.org/10.1007/JHEP09(2019)040}{{\em JHEP}
  {\bfseries 09} (2019) 040}, \href{http://arxiv.org/abs/1906.12292}{{\ttfamily
  arXiv:1906.12292 [hep-th]}}.

\bibitem{Li:2021yfk}
Y.~Li, Y.-F. Hang, H.-J. He, and S.~He, ``{Scattering amplitudes of
  Kaluza-Klein strings and extended massive double-copy},''
  \href{http://dx.doi.org/10.1007/JHEP02(2022)120}{{\em JHEP} {\bfseries 02}
  (2022) 120}, \href{http://arxiv.org/abs/2111.12042}{{\ttfamily
  arXiv:2111.12042 [hep-th]}}.

\bibitem{Gomis:2021ire}
J.~Gomis, Z.~Yan, and M.~Yu, ``{KLT factorization of winding string
  amplitudes},'' \href{http://dx.doi.org/10.1007/JHEP06(2021)057}{{\em JHEP}
  {\bfseries 06} (2021) 057}, \href{http://arxiv.org/abs/2103.05013}{{\ttfamily
  arXiv:2103.05013 [hep-th]}}.

\bibitem{Li:2022rel}
Y.~Li, Y.-F. Hang, and H.-J. He, ``{Massive color-kinematics duality and
  double-copy for Kaluza-Klein scattering amplitudes},''
  \href{http://dx.doi.org/10.1007/JHEP03(2023)254}{{\em JHEP} {\bfseries 03}
  (2023) 254}, \href{http://arxiv.org/abs/2209.11191}{{\ttfamily
  arXiv:2209.11191 [hep-th]}}.

\bibitem{Hang:2021oso}
Y.-F. Hang, H.-J. He, and C.~Shen, ``{Structure of Chern-Simons scattering
  amplitudes from topological equivalence theorem and double-copy},''
  \href{http://dx.doi.org/10.1007/JHEP01(2022)153}{{\em JHEP} {\bfseries 01}
  (2022) 153}, \href{http://arxiv.org/abs/2110.05399}{{\ttfamily
  arXiv:2110.05399 [hep-th]}}.

\bibitem{Hang:2023fkk}
Y.-F. Hang, H.-J. He, and C.~Shen, ``{Topological Equivalence Theorem and
  Double-Copy for Chern\textendash{}Simons Scattering Amplitudes},''
  \href{http://dx.doi.org/10.34133/research.0072}{{\em Research} {\bfseries 6}
  (2023) 0072}, \href{http://arxiv.org/abs/2406.13671}{{\ttfamily
  arXiv:2406.13671 [hep-th]}}.

\bibitem{Gonzalez:2021bes}
M.~C. Gonz\'alez, A.~Momeni, and J.~Rumbutis, ``{Massive double copy in three
  spacetime dimensions},''
  \href{http://dx.doi.org/10.1007/JHEP08(2021)116}{{\em JHEP} {\bfseries 08}
  (2021) 116}, \href{http://arxiv.org/abs/2107.00611}{{\ttfamily
  arXiv:2107.00611 [hep-th]}}.

\bibitem{Johnson:2020pny}
L.~A. Johnson, C.~R.~T. Jones, and S.~Paranjape, ``{Constraints on a Massive
  Double-Copy and Applications to Massive Gravity},''
  \href{http://dx.doi.org/10.1007/JHEP02(2021)148}{{\em JHEP} {\bfseries 02}
  (2021) 148}, \href{http://arxiv.org/abs/2004.12948}{{\ttfamily
  arXiv:2004.12948 [hep-th]}}.

\bibitem{Momeni:2020hmc}
A.~Momeni, J.~Rumbutis, and A.~J. Tolley, ``{Kaluza-Klein from
  colour-kinematics duality for massive fields},''
  \href{http://dx.doi.org/10.1007/JHEP08(2021)081}{{\em JHEP} {\bfseries 08}
  (2021) 081}, \href{http://arxiv.org/abs/2012.09711}{{\ttfamily
  arXiv:2012.09711 [hep-th]}}.

\bibitem{Csaki:2004ay}
C.~Csaki, ``{TASI lectures on extra dimensions and branes},'' in {\em
  {Theoretical Advanced Study Institute in Elementary Particle Physics (TASI
  2002): Particle Physics and Cosmology: The Quest for Physics Beyond the
  Standard Model(s)}}, pp.~605--698.
\newblock 4, 2004.
\newblock \href{http://arxiv.org/abs/hep-ph/0404096}{{\ttfamily
  arXiv:hep-ph/0404096}}.

\bibitem{Sundrum:2005jf}
R.~Sundrum, ``{Tasi 2004 lectures: To the fifth dimension and back},'' in {\em
  {Theoretical Advanced Study Institute in Elementary Particle Physics}:
  {Physics in D $\geqq$ 4}}, pp.~585--630.
\newblock 8, 2005.
\newblock \href{http://arxiv.org/abs/hep-th/0508134}{{\ttfamily
  arXiv:hep-th/0508134}}.

\bibitem{Rattazzi:2003ea}
R.~Rattazzi, ``{Cargese lectures on extra-dimensions},'' in {\em {Cargese
  School of Particle Physics and Cosmology: the Interface}}, pp.~461--517.
\newblock 8, 2003.
\newblock \href{http://arxiv.org/abs/hep-ph/0607055}{{\ttfamily
  arXiv:hep-ph/0607055}}.

\bibitem{Lim:2007fy}
C.~S. Lim, T.~Nagasawa, S.~Ohya, K.~Sakamoto, and M.~Sakamoto, ``{Supersymmetry
  in 5d gravity},'' \href{http://dx.doi.org/10.1103/PhysRevD.77.045020}{{\em
  Phys. Rev. D} {\bfseries 77} (2008) 045020},
  \href{http://arxiv.org/abs/0710.0170}{{\ttfamily arXiv:0710.0170 [hep-th]}}.

\bibitem{Lim:2008hi}
C.~S. Lim, T.~Nagasawa, S.~Ohya, K.~Sakamoto, and M.~Sakamoto, ``{Gauge-Fixing
  and Residual Symmetries in Gauge/Gravity Theories with Extra Dimensions},''
  \href{http://dx.doi.org/10.1103/PhysRevD.77.065009}{{\em Phys. Rev. D}
  {\bfseries 77} (2008) 065009},
  \href{http://arxiv.org/abs/0801.0845}{{\ttfamily arXiv:0801.0845 [hep-th]}}.

\bibitem{Macesanu:2005jx}
C.~Macesanu, ``{The Phenomenology of universal extra dimensions at hadron
  colliders},'' \href{http://dx.doi.org/10.1142/S0217751X06030886}{{\em Int. J.
  Mod. Phys. A} {\bfseries 21} (2006) 2259--2296},
  \href{http://arxiv.org/abs/hep-ph/0510418}{{\ttfamily arXiv:hep-ph/0510418}}.

\bibitem{Cheng:1999bg}
H.-C. Cheng, B.~A. Dobrescu, and C.~T. Hill, ``{Electroweak symmetry breaking
  and extra dimensions},''
  \href{http://dx.doi.org/10.1016/S0550-3213(00)00401-6}{{\em Nucl. Phys. B}
  {\bfseries 589} (2000) 249--268},
  \href{http://arxiv.org/abs/hep-ph/9912343}{{\ttfamily arXiv:hep-ph/9912343}}.

\bibitem{Georgi:2000wb}
H.~Georgi, A.~K. Grant, and G.~Hailu, ``{Chiral fermions, orbifolds, scalars
  and fat branes},'' \href{http://dx.doi.org/10.1103/PhysRevD.63.064027}{{\em
  Phys. Rev. D} {\bfseries 63} (2001) 064027},
  \href{http://arxiv.org/abs/hep-ph/0007350}{{\ttfamily arXiv:hep-ph/0007350}}.

\bibitem{Gherghetta:2000qt}
T.~Gherghetta and A.~Pomarol, ``{Bulk fields and supersymmetry in a slice of
  AdS},'' \href{http://dx.doi.org/10.1016/S0550-3213(00)00392-8}{{\em Nucl.
  Phys. B} {\bfseries 586} (2000) 141--162},
  \href{http://arxiv.org/abs/hep-ph/0003129}{{\ttfamily arXiv:hep-ph/0003129}}.

\bibitem{Smolyakov:2012ud}
M.~N. Smolyakov, ``{More on divergences in brane world models},''
  \href{http://dx.doi.org/10.1103/PhysRevD.87.104035}{{\em Phys. Rev. D}
  {\bfseries 87} no.~10, (2013) 104035},
  \href{http://arxiv.org/abs/1210.7978}{{\ttfamily arXiv:1210.7978 [hep-th]}}.

\bibitem{5dBCFW}
Y.~Hang, H.-J. He, and W.-W. Zhao, ``{BCFW for 4D Kaluza-Klein Theory},'' {\em
  Manuscript in preparation\hspace*{-1mm}} .

\bibitem{Britto:2004ap}
R.~Britto, F.~Cachazo, and B.~Feng, ``{New recursion relations for tree
  amplitudes of gluons},''
  \href{http://dx.doi.org/10.1016/j.nuclphysb.2005.02.030}{{\em Nucl. Phys. B}
  {\bfseries 715} (2005) 499--522},
  \href{http://arxiv.org/abs/hep-th/0412308}{{\ttfamily arXiv:hep-th/0412308}}.

\bibitem{Britto:2005fq}
R.~Britto, F.~Cachazo, B.~Feng, and E.~Witten, ``{Direct proof of tree-level
  recursion relation in Yang-Mills theory},''
  \href{http://dx.doi.org/10.1103/PhysRevLett.94.181602}{{\em Phys. Rev. Lett.}
  {\bfseries 94} (2005) 181602},
  \href{http://arxiv.org/abs/hep-th/0501052}{{\ttfamily arXiv:hep-th/0501052}}.

\end{thebibliography}\endgroup

\end{document}